\documentclass[letterpaper,twocolumn,10pt]{article}
\usepackage{usenix}
\usepackage{pifont}
\usepackage{amssymb}
\usepackage{makecell}
\usepackage{svg}
\usepackage{hyperref}
\usepackage{subfigure}
\usepackage{multirow}
\usepackage{tikz}
\usepackage{nopageno}
\usepackage{amsmath}

\newcommand{\system}{\sloppy{MAGIC\@}}

\begin{document}
\date{}

\title{\Large \bf MAGIC: Detecting Advanced Persistent Threats via Masked Graph \\ Representation Learning}

\author{\rm Zian Jia$^1$,~\rm Yun Xiong$^1$,~\rm Yuhong Nan$^2$\thanks{Corresponding author: Yuhong Nan},~\rm Yao Zhang$^1$,~\rm Jinjing Zhao$^3$,~\rm Mi Wen$^4$\\\\
$^1$Shanghai Key Laboratory of Data Science, School of Computer Science, Fudan University, China\\$^2$School of Software Engineering, Sun Yat-sen University, China\\$^3$National Key Laboratory of Science and Technology on Information System Security, China\\$^4$Shanghai University of Electric Power, China\\}
\maketitle

\begin{abstract}
Advance Persistent Threats (APTs), adopted by most delicate attackers, are becoming increasing common and pose great threat to various enterprises and institutions. Data provenance analysis on provenance graphs has emerged as a common approach in APT detection. However, previous works have exhibited several shortcomings: (1) requiring attack-containing data and \emph{a priori} knowledge of APTs, (2) failing in extracting the rich contextual information buried within provenance graphs and (3) becoming impracticable due to their prohibitive computation overhead and memory consumption. 

In this paper, we introduce \system{}, a novel and flexible self-supervised APT detection approach capable of performing multi-granularity detection under different level of supervision. \system{} leverages masked graph representation learning to model benign system entities and behaviors, performing efficient deep feature extraction and structure abstraction on provenance graphs. By ferreting out anomalous system behaviors via outlier detection methods, \system{} is able to perform both system entity level and batched log level APT detection. \system{} is specially designed to handle concept drift with a model adaption mechanism and successfully applies to universal conditions and detection scenarios. We evaluate \system{} on three widely-used datasets, including both real-world and simulated attacks. Evaluation results indicate that \system{} achieves promising detection results in all scenarios and shows enormous advantage over state-of-the-art APT detection approaches in performance overhead.
\end{abstract}

\section{Introduction}

Advanced Persistent Threats (APTs) are intentional and sophisticated cyber-attacks conducted by skilled attackers and pose great threat to both enterprises and institutions~\cite{aptsurvey-1}. Most APTs involve zero-day vulnerabilities and are especially difficult to detect due to their stealthy and changeful nature. 

Recent works~\cite{aptrulebased-2,aptrulebasedholmes-3,aptrulebased-4,aptrulebased-5,aptrulebased-6,aptstatbasednodoze-7,aptstatbased-8,aptstatbased-9,aptlearnbased-10,aptlearnbasedatlas-11,aptlearnbased-12,atplearnbased-13,aptlearnbased-14,aptlearnbasedprovgem-15,aptlearnbased-16,aptlearnbasedthreatrace-17,aptlearnbasedshadewatcher-18} on APT detection leverage data provenance to perform APT detection. Data provenance transforms audit logs into provenance graphs, which extract the rich contextual information from audit logs and provide a perfect platform for fine-grained causality analysis and APT detection. Early works~\cite{aptrulebased-2,aptrulebasedholmes-3,aptrulebased-4,aptrulebased-5,aptrulebased-6} construct rules based on typical or specific APT patterns and match audit logs against those rules to detect potential APTs. Several recent works~\cite{aptstatbasednodoze-7,aptstatbased-8,aptstatbased-9} adopt a statistical anomaly detection approach to detect APTs focusing on different provenance graph elements, e.g, system entities, interactions and communities. Most recent works~\cite{aptlearnbased-10,aptlearnbasedatlas-11,aptlearnbased-12,atplearnbased-13,aptlearnbased-14,aptlearnbasedprovgem-15,aptlearnbased-16,aptlearnbasedthreatrace-17,aptlearnbasedshadewatcher-18}, however, are deep learning-based approaches. They utilize various deep learning (DL) techniques to model APT patterns or system behaviors and perform APT detection in a classification or anomaly detection style. 

While these existing approaches have demonstrated their capability to detect APTs with reasonable accuracy, they encounter various combinations of the following challenges: (1) Supervised methods suffer from lack-of-data (LOD) problem as they require \emph{a priori} knowledge about APTs (i.e. attack patterns or attack-containing logs). In addition, these methods are particularly  vulnerable when confronted with new types of APTs they are not trained to deal with. (2) Statistics-based methods only require benign data to function, but they fail to extract the deep semantics and correlation of complex benign activities buried in audit logs, resulting in high false positive rate. (3) DL-based methods, especially sequence-based and graph-based approaches, have achieved promising effectiveness at the cost of heavy computation overhead, rendering them impractical in real-life detection scenarios.

In this paper, we address the above three issues by introducing \system{}, a novel self-supervised APT detection approach that leverages \emph{masked graph representation learning} and simple \emph{outlier detection} methods to identify key attack system entities from massive audit logs. \system{} first constructs the provenance graph from audit logs in simple yet universal steps. \system{} then employs a graph representation module that obtains embeddings by incorporating graph features and structural information in a self-supervised way. The model is built upon \emph{graph masked auto-encoders}~\cite{gmae-19} under the joint supervision of both \emph{masked feature reconstruction} and \emph{sample-based structure reconstruction}. An unsupervised outlier detection method is employed to analyze the computed embeddings and attain the final detection result. 

\system{} is designed to be flexible and scalable. Depending on the application background, \system{} is able to perform multi-granularity detection, i.e., detecting APT existence in batched logs or locating entity-level adversaries. Although \system{} is designed to perform APT detection without attack-containing data, it is well-suited for semi-supervised and fully-supervised conditions. Furthermore, \system{} also contains an optional model adaption mechanism which provides a feedback channel for its users. Such feedback is important for \system{} to further improve its performance, combat concept drift and reduce false positives. 

We implement \system{} and evaluate its performance and overhead on three different APT attack datasets: the DARPA Transparent Computing E3 datasets~\cite{darpadataset-20}, the StreamSpot dataset~\cite{streamspotdataset-21} and the Unicorn Wget dataset~\cite{unicorndataset-22}. The DARPA datasets contain real-world attacks while the StreamSpot and Unicorn Wget dataset are fully simulated in controlled environments. Evaluation results show that \system{} is able to perform entity-level APT detection with 97.26\% precision and 99.91\% recall as well as minimum overhead, less memory demanding and significantly faster than state-of-the-art approaches (e.g. 51 times faster than ShadeWatcher~\cite{aptlearnbasedshadewatcher-18}). 

To benefit future research and encourage further improvement on \system{}, we make our implementation of \system{} and our pre-processed datasets open to public\footnote{\system{} is available at \url{https://github.com/FDUDSDE/MAGIC}}. In summary, this paper makes the following contributions:

\vspace{3pt}\noindent$\bullet$~ We propose \system{}, a universal APT detection approach based on masked graph representation learning and outlier detection methods, capable of performing multi-granularity detection on massive audit logs. 

\vspace{3pt}\noindent$\bullet$~ We ensure \system{}'s practicability by minimizing its computation overhead with extended graph masked auto-encoders, allowing \system{} to complete training and detection in acceptable time even under tight conditions.

\vspace{3pt}\noindent$\bullet$~ We secure \system{}'s universality with various efforts. We leverage masked graph representation learning and outlier detection methods, enabling \system{} to perform precise detection under different supervision levels, in different detection granularity and with audit logs from various sources.

\vspace{3pt}\noindent$\bullet$~ We evaluate \system{} on three widely-used datasets, involving both real-world and simulated APT attacks. Evaluation results show that \system{} detects APTs with promising results and minimum computation overhead.

\vspace{3pt}\noindent$\bullet$~ We provide an open source implementation of \system{} to benefit future research in the community and encourage further improvement on our approach.
\section{Background}

\subsection{Motivating Example}\label{sec_motivation_1_revised}
\begin{figure}[t]
    \centering
    \includegraphics[width=0.495\textwidth]{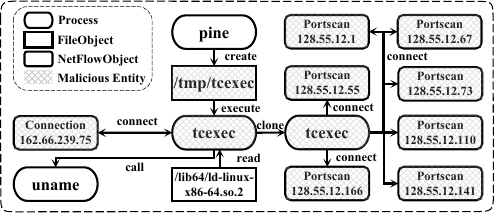}
    \caption{The provenance graph of a real-world APT attack, exploiting the Pine Backdoor vulnerability. All attack-irrelevant entities and interactions have been removed from the provenance graph.}
    \label{fig_motivation_revised}
  \end{figure}

Here we provide a detailed illustration of an APT scenario that we use throughout the paper. \emph{Pine backdoor with Drakon Dropper} is an APT attack from the DARPA Engagement 3 Trace dataset~\cite{darpadataset-20}. During the attack, an attacker constructs a malicious executable (\emph{/tmp/tcexec}) and sends it to the target host via a phishing e-mail. The user then unconsciously downloads and opens the e-mail. Contained within the e-mail is an executable designed to perform a port-scan for internal reconnaissance and establish a silent connection between the target host and the attacker. Figure~\ref{fig_motivation_revised} displays the provenance graph of our motivation example. Nodes in the graph represent system entities and arrows represent directed interactions between entities. The graph shown is a subgraph abstracted from the complete provenance graph by removing most attack-irrelevant entities and interactions. Different node shape corresponds to different type of entities. Entities covered in stripes are considered malicious ones.

\subsection{Prior Research and their Limitations}\label{sec_motivation_2_revised}

\noindent\textbf{Supervised Methods.} For early works~\cite{aptrulebased-2,aptrulebased-4,aptrulebasedholmes-3,aptrulebased-5,aptrulebased-6}, special heuristic rules need to be constructed to cover all attack patterns. Many DL-based APT detection methods~\cite{aptlearnbasedatlas-11,aptlearnbased-14,aptlearnbasedprovgem-15,aptlearnbased-16,aptlearnbasedshadewatcher-18} construct provenance graphs based on both benign and attack-containing data and detect APTs in a classification style. These supervised methods can achieve almost perfect detection results on learned attack patterns but are especially vulnerable facing \emph{concept drift} or unseen attack patterns. Moreover, for rule-based methods, the construction and maintenance of heuristic rules can be very expensive and time-consuming. And for DL-based methods, the scarcity of attack-containing data is preventing these supervised methods from being actually deployable. To address the above issue, \system{} adopts a fully self-supervised anomaly detection style, allowing the absence of attack-containing data while effectively dealing with unseen attack patterns. 

\noindent\textbf{Statistics-based Methods.} Most recent statistics-based methods~\cite{aptstatbasednodoze-7,aptstatbased-8,aptstatbased-9} detect APT signals by identifying system entities, interactions and communities based on their rarity or anomaly scores. However, the rarity of system entities may not necessarily indicate their abnormality and anomaly scores, obtained via causal analysis or label propagation, are \emph{shallow feature extraction} on provenance graphs. To illustrate, the process \emph{tcexec} performs multiple portscan operations on different IP addresses in our motivating example (See Figure~\ref{fig_motivation_revised}), which may be considered as a normal system behavior. However, taking into consideration that process \emph{tcexec}, derived from the external network, also reads sensitive system information (\emph{uname}) and makes connection with public IP addresses (\emph{162.66.239.75}), we can easily identify \emph{tcexec} as a malicious entity. Failure to extract deep semantics and correlations between system entities often results in low detection performance and high false positive rate of statistics-based methods. \system{}, however, employs a graph representation module to perform \emph{deep graph feature extraction} on provenance graphs, resulting in high-quality embeddings.

\noindent\textbf{DL-based Methods.} Recently, DL-based APT detection methods, no matter supervised or unsupervised, are producing very promising detection results. However, in reality, hundreds of GB of audit logs are produced every day in a medium-size enterprise~\cite{datascale-23}. Consequently, DL-based methods, especially sequence-based~\cite{aptlearnbasedatlas-11,aptlearnbased-14, airtag-60} and graph-based~\cite{aptlearnbased-10,aptlearnbased-12,aptlearnbasedprovgem-15,aptlearnbased-16,aptlearnbasedthreatrace-17,aptlearnbasedshadewatcher-18} methods, are impracticable due to their computation overhead. For instance, ATLAS~\cite{aptlearnbasedatlas-11} takes an average 1 hour to train on 676MB of audit logs and ShadeWatcher~\cite{aptlearnbasedshadewatcher-18} takes 1 day to train on the DARPA E3 Trace dataset with GPU available. Besides, some graph auto-encoder~\cite{vgae-24,gala-25,gate-26} based methods encounter \emph{explosive memory overhead} problem when the scale of provenance graphs expands. \system{} avoids to be computationally demanding by introducing \emph{graph masked auto-encoders} and completes its training on the DARPA E3 Trace dataset in mere minutes. Detailed evaluation of \system{}'s performance overhead is presented in Sec.~\ref{sec_evaluation_4_revised}.

\noindent\textbf{End-to-end Approaches}. Beyond the three major limitations discussed above, it is also worth to mention that most recent APT detection approaches~\cite{aptlearnbasedthreatrace-17,aptlearnbasedshadewatcher-18,aptlearnbasedatlas-11} are \emph{end-to-end} detectors and focus on one specific detection task. For instance, ATLAS~\cite{aptlearnbasedatlas-11} focused on end-to-end attack reconstruction and Unicorn~\cite{aptlearnbased-10} yields system-level alarms from streaming logs. Instead, \system{}'s approach is universal and performs multi-granularity APT detection under various detection scenarios, which can also be applied to audit logs collected from different sources. 

\subsection{Threat Model and Definitions}\label{sec_definition_revised}
We first present the threat model we use throughout the paper and then formally define key concepts that are crucial to understanding how \system{} performs APT detection.

\noindent\textbf{Threat Model.} We assume that attackers come from outside a system and target valuable information within the system. An attacker may perform sophisticated steps to achieve his goal but leaves trackable evidence in logs. The combination of the system hardware, operating system and system audit softwares is our trusted computing base. Poison attacks and evasion attacks are not considered in our threat model.

\noindent\textbf{Provenance Graph.} A provenance graph is a directed cyclic graph extracted from raw audit logs. Constructing a provenance graph is common practice in data provenance, as it connects system entities and presents the interaction relationships between them. A provenance graph contains nodes representing different system entities (e.g., processes, files and sockets) and edges representing interactions between system entities (e.g., execute and connect), labeled with their types. For example, \emph{/tmp/tcexec} is a \emph{FileObject} system entity and the edge between \emph{/tmp/tcexec} and \emph{tcexec} is an \emph{execute} operation from a \emph{FileObject} targeting a \emph{Process} (See Figure~\ref{fig_motivation_revised}).

\noindent\textbf{Multi-granularity Detection.} \system{} is capable to perform APT detection at two-granularity: \emph{batched log level} and \emph{system entity level}. \system{}'s multi-granularity detection ability gives rises to a two-stage detection approach: first conduct batched log level detection on streaming batches of logs, and then perform system entity level detection on positive batches to identify detailed detection results. Applying this approach to real-world settings will effectively reduce workload, resource consumption and false positives and, in the meantime, produce detailed outcomes.

\vspace{3pt}\noindent$\bullet$~ Batched log level detection. Under this granularity of APT detection, the major task is \emph{given batched audit logs from a consistent source, \system{} alerts if a potential APT is detected in a batch of logs}. Similar to Unicorn~\cite{aptlearnbased-10}, \system{} does not accurately locate malicious system entities and interactions under this granularity of detection.

\vspace{3pt}\noindent$\bullet$~ System entity level detection. The detection task under this granularity of APT detection is \emph{given audit logs from a consistent source, \system{} is able to accurately locate malicious system entities in those audit logs}. Identification of key system entities during APTs is vital to subsequent tasks such as attack investigation and attack story recovery as it provides explicable detection results and reduces the need for domain experts as well as redundant manual efforts~\cite{aptlearnbasedatlas-11}.
\begin{figure*}[t]
    \centering
    \includegraphics[width=0.99\hsize]{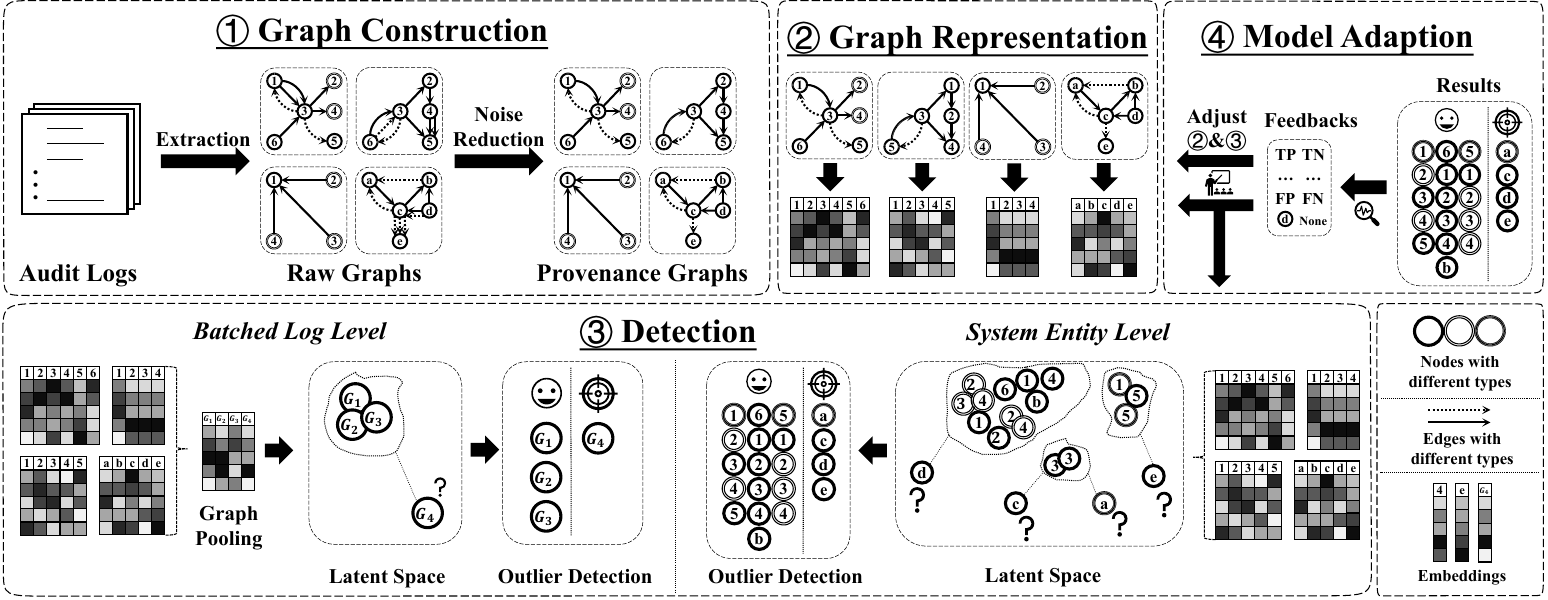}
    \caption{Overview of \system{}'s detection pipeline.}
    \label{fig_overview_revised}
  \end{figure*}
\section{\system{} Overview}\label{sec_overview_revised}

\system{} is a novel self-supervised APT detection approach that leverages masked graph representation learning and outlier detection methods and is capable of efficiently performing multi-granularity detection on massive audit logs. \system{}'s pipeline consists of three main components: (1) provenance graph construction, (2) a graph representation module and (3) a detection module. It also provides an optional (4) model adaption mechanism. During training, \system{} transforms training data with (1), learns graph embedding by (2) and memorizes benign behaviors in (3). During inference, \system{} transforms target data with (1), obtains graph embedding with the trained (2) and detects outliers through (3). Figure~\ref{fig_overview_revised} gives an overview of the \system{} architecture.

Streaming audit logs collected by system auditing softwares are usually stored in batches. During provenance graph construction (1), \system{} transforms these logs into static provenance graphs. System entities and interactions between them are extracted and converted into nodes and edges respectively. Several complexity reduction techniques are utilized to remove redundant information.

The constructed provenance graphs are then fed through the graph representation module (2) to obtain output embeddings (i.e. comprehensive vector representations of objects). Built upon \emph{graph masked auto-encoders} and integrating \emph{sample-based structure reconstruction}, the graph representation module embeds, propagates and aggregates node and edge attributes into output embeddings, which contain both node embeddings and the system state embedding. 

The graph representation module is trained with only benign audit logs to model benign system behaviors. When performing APT detection on potentially attack-containing audit logs, \system{} utilizes \emph{outlier detection methods} based on the output embeddings to detect outliers in system behaviors (3). Depending on the granularity of the task, different embeddings are used to complete APT detection. On \emph{batched log level} tasks, the system state embeddings, which reflect the general behaviors of the whole system, are the detection targets. An outlier in such embeddings means its corresponding system state is unseen and potentially malicious, which reveals an APT signal in that batch. On \emph{system entity level} tasks, the detection targets are those node embeddings, which represent the behaviors of system entities. Outliers in node embeddings indicates suspicious system entities and detects APT threats in finer granularity.

In real-world detection settings, \system{} has two pre-designed applications. For each batch of logs collected by system auditing softwares, one can either directly utilize \system{}'s entity level detection to accurately identify malicious entities within the batch, or perform a two-stage detection, as stated in Sec.~\ref{sec_definition_revised}. In this case, \system{} first scans a batch and sees if malicious signals exist in the batch (batched log level detection). If it alerts positive, \system{} then performs entity level detection to identify malicious system entities in finer granularity. Batched log level detection is significantly less computationally demanding than entity level detection. Therefore, such a two-stage routine can help \system{}'s users to save computational resource and avoid false alarms without affecting \system{}'s detection fineness. However, if users favor fine-grain detection on all system entities, the former routine is still an accessible option.

To deal with \emph{concept drift} and unseen attacks, an optional model adaption mechanism is employed to provide feedback channels for its users (4). Detection results checked and confirmed by security analysts are fed back to \system{}, helping it to adapt to benign system behavior changes in a semi-supervised way. Under such conditions, \system{} achieves even more promising detection results, which is discussed in Sec.~\ref{sec_evaluation_3_revised}. Furthermore, \system{} can be easily applied to real-world online APT detection thanks to it's ability to adapt itself to \emph{concept drift} and its minimum computation overhead.

\section{Design Details}\label{sec_model_revised}
In this section, we explain in detail how \system{} performs efficient APT detection on massive audit logs. \system{} contains four major components: a graph construction phase that builds optimised and consistent provenance graphs (Sec.~\ref{sec_model_1_revised}), a graph representation module that produces output embeddings with maximum efficiency (Sec.~\ref{sec_model_2_revised}), a detection module that utilizes outlier detection methods to perform APT detection (Sec.~\ref{sec_model_3_revised}) and a model adaption mechanism to deal with \emph{concept drift} and other high-quality feedbacks (Sec.~\ref{sec_model_4_revised}).

\subsection{Provenance Graph Construction}\label{sec_model_1_revised}

\system{} first constructs a provenance graph out of raw audit logs before performing graph representation and APT detection. We follow three steps to construct a consistent and optimised provenance graph ready for graph representation.

\begin{figure}[t]
    \centering
    \includegraphics[width=0.46\textwidth]{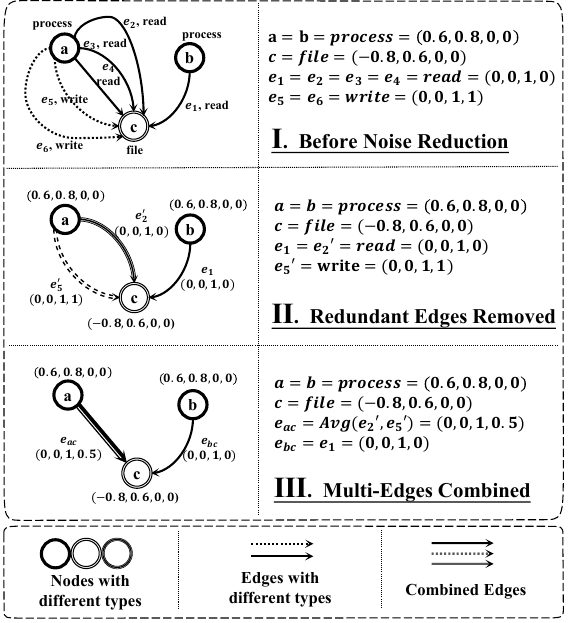}
    \caption{Example of \system{}'s graph construction steps.}
    \label{fig_gc_revised}
  \end{figure}

\noindent\textbf{Log Parsing.} The first step is to simply parse each log entry, extract system entities and system interactions between them. Then, a prototype provenance graph can be constructed with system entities as nodes and interactions as edges. Now we extract categorical information regarding nodes and edges. For simple log format that provides entity and interaction labels, we directly utilize these labels. For some format that provides complicated attributes of those entities and interactions, we apply multi-label hashing (e.g., xxhash~\cite{xxhash-27}) to transform attributes into labels. At this stage, the provenance graph is a directed multi-graph. We designed an example to demonstrate how we deal with the raw provenance graph after log parsing in Figure~\ref{fig_gc_revised}.

\noindent\textbf{Initial Embedding.} In this stage, we transform node and edge labels into fixed-size feature vector (i.e., initial embedding) of dimension $d$, where $d$ is the hidden dimension of our graph representation module. We apply a \emph{lookup embedding}, which establish an one-to-one mapping between node/edge labels to $d$-dimension feature vectors. As demonstrated in Figure~\ref{fig_gc_revised} (I and II), process \emph{a} and \emph{b} share the same label, so they are mapped to the same feature vector, while \emph{a} and \emph{c} are embedded into different feature vectors as they have different labels. We note that the possible number of unique node/edge labels is determined by the data source (i.e., auditing log format). Therefore, the lookup embedding works under a transductive setting and do not need to learn embeddings for unseen labels.

\noindent\textbf{Noise Reduction.} The expected input provenance graph of our graph representation module would be simple-graphs. Thus, we need to combine multiple edges between node pairs. If multiple edges of the same label (also sharing the same initial embedding) exist between a pair of nodes, we remove redundant edges so that only one of them remains. Then we combine the remaining edges into one final edge. We note that between a pair of nodes, edges of several different labels may remain. After the combination, the initial embedding of the resulting unique edge is obtained by averaging the initial embeddings of the remaining edges. To illustrate, we show how our noise reduction combines multi-edges and how it affects the edge initial embeddings in Figure~\ref{fig_gc_revised} (II an III). First, three \emph{read} and two \emph{write} interactions between \emph{a} and \emph{c} are merged into one for each label. Then we combine them together, forming one edge $e_{ac}$ with initial embedding equal to the average initial embedding of the remaining edges ($e_2^{\prime}$ and $e_5^{\prime}$). We provide a comparison between our noise reduction steps and previous works in Appendix~\ref{appe_c_revised}.

After conducting the above three steps, \system{} has finished constructing a consistent and information-preserving provenance graph ready for subsequent tasks. During provenance graph construction, little information is lost as \system{} only damages the original semantics by generalizing detailed descriptions of system entities and interactions into labels. However, an average 79.60\% of all edges are reduced on the DARPA E3 Trace dataset, saving \system{}'s training time and memory consumption.

\subsection{Graph Representation Module}\label{sec_model_2_revised}

\begin{figure}[t]
    \centering
    \includegraphics[width=0.47\textwidth]{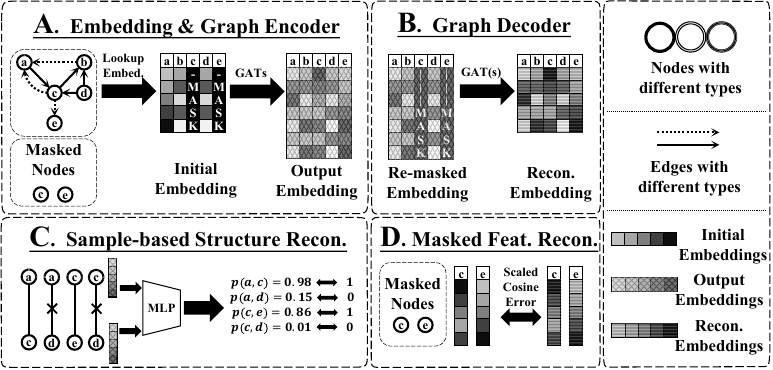}
    \caption{Graph representation module of \system{}.}
    \label{fig_gmae_revised}
  \end{figure}
\system{} employs a graph representation module to obtain high-quality embeddings from featured provenance graphs. As illustrated in Figure~\ref{fig_gmae_revised}, the graph representation module consists of three phases: a masking procedure to partially hide node features (i.e. initial embeddings) for reconstruction purpose (Sec.~\ref{sec_model_2_1_revised}), a graph encoder that produces node and system state output embeddings by propagating and aggregating graph features (Sec.~\ref{sec_model_2_2_revised}), a graph decoder that provides supervision signals for the training of the graph representation module via masked feature reconstruction and sample-based structure reconstruction (Sec.~\ref{sec_model_2_3_revised}). The encoder and decoder form a \emph{graph masked auto-encoder}, which excels in producing fast and resource-saving embeddings.

\subsubsection{Feature Masking}\label{sec_model_2_1_revised}

Before training our graph representation module, we perform \emph{masking} on nodes, so that the graph masked auto-encoder can be trained upon reconstruction of these nodes. Masked nodes are randomly chosen, covering a certain proportion of all nodes. The initial embeddings of such masked nodes are replaced with a special mask token $x_{mask}$ to cover any original information about these nodes. Edges, however, are not masked because these edges provide precious information about relationships between system entities. In summary, given node initial embeddings $x_n$, we mask nodes as follows:
$$
\begin{aligned}
\begin{split}
emb_{n} &= 
\begin{array}{ll}
x_n, & {n \notin \widetilde{N}}\\
x_{mask}, & {n \in \widetilde{N}}
\end{array}
\end{split}
\end{aligned}
$$
where $\widetilde{N}$ are randomly-chosen masked nodes, $emb_n$ is the embedding of node $n$ ready for training the graph representation module. This masking process only happens during training. During detection, we do not mask any nodes.

\subsubsection{Graph Encoder}\label{sec_model_2_2_revised}

Initial embeddings obtained from the graph construction steps take only raw features into consideration. However, raw features are far from enough to model detailed behaviors of system entities. Contextual information of an entity, such as its neighborhood, its multi-hop relationships and its interaction patterns with other system entities plays an important role to obtain high-quality entity embeddings~\cite{gin-28}. Here we employ and extend graph masked auto-encoders~\cite{gmae-19} to generate output embeddings in a self-supervised way. The graph masked auto-encoder consists of an encoder and a decoder. The encoder produces output embeddings by propagating and aggregating graph features and the decoder reconstructs graph features to provide supervision signals for training. Such encoder-decoder architecture maintains the contextual and semantic information within the generated embeddings, while its computation overhead is significantly reduced via masked learning.

The encoder of our graph representation module contains multiple stacked layers of graph attention networks (GAT)~\cite{gat-29}. The function of a GAT layer is to generate output node embeddings according to both the features (initial embeddings) of the node itself and its neighbors. Differing from ordinary GNNs, GAT introduces an attention mechanism to measure the importance of those neighbors.

To explain in detail, one layer of GAT takes node embeddings generated by previous layers as input and propagates embeddings from source nodes to destination nodes into messages along the interactions. The message contains information about the source node and the interaction between source and destination:
$$
MSG(src, dst) = W_{msg}^T(h_{src}||emb_e).\\
$$

And the attention mechanism is employed to calculate the attention coefficients between the message source and its destination:
$$
\begin{aligned}
\alpha(src, dst) &= LeakyReLU(W_{as}^Th_{src}+W_{am}MSG(src, dst)),\\
a(src, dst) &= Softmax(\alpha(src, dst)).\\
\end{aligned}
$$

Then for the destination node, the GAT aggregates messages from incoming edges to update its node embedding by computing a weighted sum of all incoming messages. The weights are exactly the attention coefficients:
$$
\begin{aligned}
AGG(h_{dst}, h_{\mathcal{N}}) &= W_{self}h_{dst} + \sum_{i\in \mathcal{N}}a(i, dst)MSG(i, dst),\\
h_n^l &= AGG^l(h_n^{l-1}, h_{\mathcal{N}_n}^{l-1}).
\end{aligned}
$$
where $h_n^l$ is the hidden embedding of node $n$ at $l$-th layer of GAT, $h_n^{l-1}$ is that of layer $l-1$ and $\mathcal{N}_n$ is the one-hop neighborhood of $n$. The input of the first GAT layer are the initial node embeddings. $emb_e$ is the initial edge embedding and remains constant throughout the graph representation module. $W_{as},W_{am},W_{self},W_{msg}$ are trainable parameters. The updated node embedding forms a general abstraction of the node's one-hop interaction behavior.

Multiple layers of such GATs are stacked to obtain the final node embedding $h$, which is concatenated by the original node embedding and outputs of all GAT layers:
$$
\begin{aligned}
h_n = emb_n || h_n^1 || \cdots || h_n^l.
\end{aligned}
$$
where $\cdot||\cdot$ denotes the concatenate operation. The more layers of GAT stacked, the wider the neighboring range is and the farther a node's multi-hop interaction pattern its embedding is able to represent. Consequently, the graph encoder effectively incorporates node initial features and multi-hop interaction behaviors to abstract system entity behaviors into node embeddings. The graph encoder also applies an average pooling to all node embeddings to generate a comprehensive embedding of the graph itself~\cite{graphpooling-30}, which recapitulates the overall state of the system:
$$
h_G = \frac{1}{|N|}\sum_{n_i\in N}h_{n_i}.
$$

The node embeddings and system state embeddings generated by the graph encoder are considered the output of the graph representation module, which are used in subsequent tasks in different scenarios.
\subsubsection{Graph Decoder}\label{sec_model_2_3_revised}
The graph encoder does not provide supervision signals that support model training. In typical graph auto-encoders~\cite{vgae-24,gate-26}, a graph decoder is employed to decode node embeddings and supervise model training via \emph{feature reconstruction} and \emph{structure reconstruction}. Graph masked auto-encoders, however, abandon structure reconstruction to reduce computation overhead. Our graph decoder is a mixture of both, which integrates masked feature reconstruction and sample-based structure reconstruction to construct an objective function that optimizes the graph representation module.

Given node embeddings $h_n$ obtained from the graph encoder, the decoder first re-masks those masked nodes and transforms them into the input of \emph{masked feature reconstruction}:
$$
\begin{aligned}
\begin{split}
h^*_n &= \left\{
\begin{array}{ll}
W^*h_n, & {n \notin \widetilde{N}}\\
W^*v_{remask}, & {n \in \widetilde{N}}
\end{array}
\right.,\\
\end{split}
\end{aligned}
$$
Subsequently, the decoder uses a similar GAT layer described above to reconstruct the initial embeddings of the masked nodes, allowing the calculation of a feature reconstruction loss:
$$
\begin{aligned}
x^*_n &= AGG^*(h^*_n, h^*_{\mathcal{N}_n}),\\
L_{fr} &= \frac{1}{|\widetilde{N}|}\sum_{n_i\in \widetilde{N}}(1-\frac{x_{n_i}^Tx^*_{n_i}}{||x_{n_i}||\cdot||x^*_{n_i}||})^\gamma.
\end{aligned}
$$
where $L_{fr}$ is the masked feature reconstruction loss obtained by calculating a \emph{scaled cosine loss} between initial and reconstructed embeddings of the masked nodes. This loss~\cite{gmae-19} scales dramatically between easy and difficult samples which effectively speeds up learning. The degree of such scaling is controlled by a hyper-parameter $\gamma$.

Meanwhile, sample-based structure reconstruction aims to reconstruct graph structure (i.e. predict edges between nodes). Instead of reconstructing the whole adjacency matrix, which has $O(N^2)$ complexity, sample-based structure reconstruction applies contrastive sampling on node pairs and predicts edge probabilities between such pairs. Only non-masked nodes are involved in structure reconstruction. Positive samples are constructed with all existing edges between non-masked nodes and negative samples are sampled among node pairs with no existing edges between them.

A simple two-layer MLP is used to reconstruct edges between node pairs samples, generating one probability for each sample. The reconstruction loss takes the form of a simple binary cross-entropy loss on those samples:
$$
\begin{aligned}
prob(n, n') &= \sigma(MLP(h_n||h_{n'})),\\
L_{sr} = -\frac{1}{|\hat{N}|}\sum_{n\in \hat{N}}(&log(1-prob(n,n^-))+log(prob(n,n^+))).
\end{aligned}
$$
where $(n,n^-)$ and $(n,n^+)$ are negative and positive samples respectively and $\hat{N}=N - \widetilde{N}$ is the set of non-masked nodes. Sample-based structure reconstruction only provides supervision to the output embeddings. Instead of using dot products, we employ a MLP to calculate edge probabilities as interacting entities are not necessarily similar in behaviors. Also, we are not forcing the model to learn to predict edge probabilities. The function of such structure reconstruction is to \emph{maximize behavioral information} contained in the abstracted node embeddings so that a simple MLP is sufficient to incorporate and interpret such information into edge probabilities.

The final objective function $L = L_{fr} + L_{sr}$ combines $L_{fr}$ and $L_{sr}$ and provides supervision signals to the graph representation module, enabling it to learn parameters in a self-supervised way.
\subsection{Detection Module}\label{sec_model_3_revised}

Based on the output embeddings generated by the graph representation module, we utilize outlier detection methods to perform APT detection in an unsupervised way. As detailedly explained in previous sections, such embeddings summarize system behaviors in different granularity. The goal of our detection model is to identify malicious system entities or states given only a priori knowledge of \emph{benign system behaviors}. Embeddings generated via graph representation learning tend to form clusters if their corresponding entities share similar interaction behaviors in the graph~\cite{gate-26, gala-25,vgae-24,gmae-19,heco-32}. Thus, outliers in system state embeddings indicate uncommon and suspicious system behaviors. Based on such insight, we develop a special outlier detection method to perform APT detection. 

During training, benign output embeddings are first abstracted from the training provenance graphs. What the detection module does at this stage is simply \emph{memorizing} those embeddings and organize them in a \emph{K-D Tree}~\cite{kdtree-33}. After training, the detection module reveals outliers in three steps: k-nearest neighbor searching, similarity calculation and filtering. Given a target embedding, the detection module first obtains its k-nearest neighbors via K-D Tree searching. Such searching process only takes $log(N)$ time, where $N$ is the total number of memorized training embeddings. Then, a similarity criterion is applied to evaluate the target embedding's closeness to its neighbors and compute an anomaly score. If its anomaly score yields higher than a hyper-parameter $\theta$, the target embedding is considered an outlier and its corresponding system entity or system state is malicious. An example workflow of the detection module is formalized as follows, using euclidean distance as the similarity criterion:
$$
\begin{aligned}
\begin{split}
\mathcal{N}_x &= KNN(x)\\
dist_x &= \frac{1}{|\mathcal{N}_x|}\sum_{x_i\in \mathcal{N}_x}||x-x_i||\\
score_x &= \frac{dist_x}{\overline{dist}}\\
result_x &= \left\{
\begin{array}{ll}
1, & score_x\geq\theta\\
0, & score_x<\theta
\end{array}
\right.\\
\end{split}
\end{aligned}
$$
where $\overline{dist}$ is the average distance between training embeddings and their k-nearest neighbors. When performing \emph{batched log level detection}, the detection module memorizes benign system state embeddings that reflects system states and detects if the system state embedding of a newly-arrived provenance graph is an outlier. When performing \emph{system entity level detection}, the detection module instead memorizes benign \emph{node embeddings} that indicates system entity behaviors and given a newly-arrived provenance graph, it detects outliers within the embeddings of all system entities. 

\subsection{Model Adaption}\label{sec_model_4_revised}
For an APT detector to effectively function in real-world detection scenarios, concept drift must be taken into consideration. When facing benign yet previously unseen system behaviors, \system{} produces false positive detection results, which may mislead subsequent applications (e.g. attack investigation and story recovery). Recent works address this issue by forgetting outdated data~\cite{aptlearnbased-10} or fitting their model to benign system changes via a \emph{model adaption mechanism}~\cite{aptlearnbasedshadewatcher-18}. \system{} also integrates a model adaption mechanism to combat concept drift and learn from false positives identified by security analysts. Slightly different from other works that use only false positives to retrain the model, \system{} can be retrained with all feedbacks. As discussed in previous sections, the graph representation module in \system{} encodes system entities into embeddings in a self-supervised way, without knowing its label. Any unseen data, including those true negatives, are valuable training data for the graph representation module to enhance its representation ability on unseen system behaviors. 

The detection module can only be retrained with benign feedbacks to keep up to system behavior changes. And as it memorizes more and more benign feedbacks, its detection efficiency is lowered. To address this issue, we also implement a discounting mechanism on the detection module. When the volume of memorized embeddings exceeds a certain amount, earliest embeddings are simply removed as newly-arrived embeddings are learned. We provide the model adaption mechanism as an optional solution to concept drift and unseen system behaviors. It is recommended to adapt \system{} to system changes by feeding confirmed false positive samples to \system{}'s model adaption mechanism.

\section{Implementation}\label{sec_implementation_revised}

We implement \system{} with about 3,500 lines of code in Python 3.8. We develop several log parsers to cope with different format of audit logs, including StreamSpot~\cite{streamspot-35}, Camflow~\cite{camflow-34} and CDM~\cite{cdm-36}. Provenance graphs are constructed using the graph processing library Networkx~\cite{networkx-37} and stored in JSON format. The graph representation module is implemented via PyTorch~\cite{pytorch-38} and DGL~\cite{dgl-40}. The detection module is developed with Scikit-learn~\cite{sklearn-39}. For hyper-parameters of MAGIC, the scaling factor $\gamma$ in the feature reconstruction loss is set to 3, the number of neighbors k is set to 10, the learning rate as 0.001 and the weight decay factor equals $5\times10^{-4}$. We use a 3 layer graph encoder and a mask rate of 0.5 in our experiments. The output embedding dimension $d$ is different on two detection scenarios, batched log level detection and entity level detection. We use $d$ equals 256 in batched log level detection and of and an we set $d$ equals 64 in entity level detection to reduce resource consumption. The detection threshold $\theta$ is chosen by a simple linear search separately conducted on each dataset. The hyper-parameters may have other choices. We demonstrate the impact of these hyper-parameters on MAGIC later in the evaluation section. In our hyper-parameter analysis, $d$ is chosen from \{16, 32, 64, 128, 256\} , $l$ from \{1, 2, 3, 4\} and $r$ from \{0.3, 0.5, 0.7\}. For the threshold $\theta$, it is chosen between 1 and 10 in batched log level detection. For entity level detection, please refer to Appendix \ref{appe_f_revised}.
\section{Evaluation}\label{sec_evaluation_revised}

We use 131GB of audit logs derived from various system auditing softwares to evaluate the effectiveness and efficiency of \system{}. We first describe our experimental settings (Sec.~\ref{sec_evaluation_1_revised}), then elaborate the effectiveness of \system{} in different scenarios (Sec.~\ref{sec_evaluation_2_revised}), conduct a false positive analysis and assess the usefulness of the model adaption mechanism (Sec.~\ref{sec_evaluation_3_revised}) and analyze the run-time performance overhead of \system{} (Sec.~\ref{sec_evaluation_4_revised}). The impact of different components and hyper-parameters of \system{} is analyzed in Sec.~\ref{sec_evaluation_5_revised}. In addition, a detailed case study on our motivation example is conducted in Appendix~\ref{sec_evaluation_6_revised} to illustrate how \system{}'s pipeline work for APT detection. These experiments are conducted under the same device setting.

\subsection{Experimental Settings}\label{sec_evaluation_1_revised}
We evaluate the effectiveness of \system{} on three public datasets: the StreamSpot dataset~\cite{streamspotdataset-21}, the Unicorn Wget dataset~\cite{unicorndataset-22} and the DARPA Engagement 3 datasets~\cite{darpadataset-20}. These datasets vary in volume, origin and granularity. We believe by testing \system{}'s performance on these datasets, we are able to compare \system{} with as many \emph{state-of-the-art} APT detection approaches as possible and explore the universality and applicability of \system{}. we provide detailed descriptions of the three datasets as follows.

\noindent\textbf{StreamSpot Dataset.} The StreamSpot dataset (See Table~\ref{tab_batched_dataset}) is a simulated dataset collected and made public by StreamSpot~\cite{streamspot-35} using auditing system SystemTap~\cite{systemtap-41}. The StreamSpot dataset contains 600 batches of audit logs monitoring system calls under 6 unique scenarios. Five of those scenarios are simulated benign user behaviors while the attack scenario simulates a drive-by-download attack. The dataset is considered a relatively small dataset and since no labels of log entries and system entities are provided, we perform \emph{batched log level detection} on the StreamSpot dataset similar to previous works~\cite{aptlearnbased-10,aptlearnbasedprovgem-15,aptlearnbasedthreatrace-17}.

\begin{table}[t]
\caption{Datasets for batched log level detection.}
\label{tab_batched_dataset}
\resizebox{\columnwidth}{!}{
\begin{tabular}{|c|c|c|c|c|c|c|}
\hline
Dataset & Scenario & Malicious & \#Log pieces & Avg. \#Entity & Avg. \#Interaction & Size(GB)\\
\hline\hline
 \multirow{6}{*}{StreamSpot} & CNN & & 100 & 8,989 & 294,903 & 0.9\\

 & Download & & 100 & 8,830 & 310,814 & 1.0\\

 & Gmail & & 100 & 6,826 & 37,382 & 0.1\\

 & VGame & & 100 & 8,636 & 112,958 & 0.4\\

 & YouTube & & 100 & 8,292 & 113,229 & 0.3\\

 & Attack & \checkmark & 100 & 8,890 & 28,423 & 0.1\\
 \hline\hline
 
 \multirow{2}{*}{Unicorn Wget} & Benign & & 125 & 265,424 & 975,226 & 64.0\\

 & Attack & \checkmark & 25 & 257,156 & 949,887 & 12.6\\
  \hline
\end{tabular}
}
\end{table}

\noindent\textbf{Unicorn Wget Dataset.} The Unicorn Wget dataset (See Table~\ref{tab_batched_dataset}) contains simulated attacks designed by Unicorn~\cite{aptlearnbased-10}. Specifically, it contains 150 batches of logs collected with Camflow~\cite{camflow-34}, where 125 of them are benign and 25 of them contain \emph{supply-chain attacks}. Those attacks, categorized as stealth attacks, are elaborately designed to behave similar to benign system workflows and are expected to be difficult to identify. This dataset is considered the hardest among our experimental datasets for its huge volume, complicated log format and the stealthy nature of these attacks. The same as state-of-the-art approaches, we perform \emph{batched log level detection} on this dataset.

\noindent\textbf{DARPA E3 Datasets.} The DARPA Engagement 3 datasets (See Table~\ref{tab_entity_dataset_revised}), as a part of the DARPA Transparent Computing program, are collected among an enterprise network during an adversarial engagement. APT attacks exploiting different vulnerabilities~\cite{darpadataset-20} are conducted by the red team to exfiltrate sensitive information. Blue teams try to identify those attacks by auditing the network hosts and performing causality analysis on them. Trace, THEIA and CADETS sub-datasets are included in our evaluation. These three sub-datasets consist of a total 51.69GB of audit records, containing as many as 6,539,677 system entities and 68,127,444 interactions. Thus, we evaluate \system{}'s \emph{system entity level detection} ability and address the overhead issue on these datasets.

\begin{table}[t]
\caption{Datasets for system entity level detection.}
\label{tab_entity_dataset_revised}
\resizebox{\columnwidth}{!}{
\begin{tabular}{|c|c|c|c|c|c|c|}
\hline
Dataset & Scenario & Malicious &\#Node &\#Edge & Size (GB)\\
\hline\hline
 \multirow{4}{*}{DARPA E3 Trace} & Benign & & 3,220,594 & \multirow{4}{*}{4,080,457} & \multirow{4}{*}{15.40}\\

 & Extension Backdoor & \checkmark & 732 & & \\

 & Pine Backdoor & \checkmark & 67,345 & & \\

 & Phishing Executable & \checkmark & 5 & & \\
 \hline\hline
 \multirow{2}{*}{DARPA E3 THEIA} & Benign & & 1,598,647 & \multirow{2}{*}{2,874,821} & \multirow{2}{*}{17.91}\\

 & Attack & \checkmark & 25,319 & & \\
\hline\hline
\multirow{2}{*}{DARPA E3 CADETS} & Benign & & 1,614,189 & \multirow{2}{*}{3,303,264} & \multirow{2}{*}{18.38}\\

 & Attack & \checkmark & 12,846 & & \\
\hline
\end{tabular}
}
\end{table}

For different dataset, we employ different dataset splits to evaluate the model and we use only benign samples for training. For the StreamSpot dataset, we randomly choose 400 batches out of 500 benign logs for training and the rest for testing, resulting in an balanced test set. For the Unicorn Wget dataset, 100 batches of benign logs are selected for training while the rest are for testing. For the DARPA E3 datasets, we use the same ground-truth labels as ThreaTrace~\cite{aptlearnbasedthreatrace-17} and split log entries according to their order of occurrence. The earliest 80\% log entries are for training while the rest are preserved for testing. During evaluation, the average performance of \system{} under 100 global random seeds is reported as the final result, so the experimental results may contain fractions of system entities/batches of logs.

\subsection{Effectiveness}\label{sec_evaluation_2_revised}
\system{}'s effectiveness of multi-granularity APT detection is evaluated on three datasets. Here we present the detection results of \system{} on each dataset, then compare it with state-of-the-art APT detection approaches on those datasets.

\begin{figure}[t]
    \centering
    \subfigure[ROC curves on batched log level detection]{\includegraphics[width=0.495\hsize]{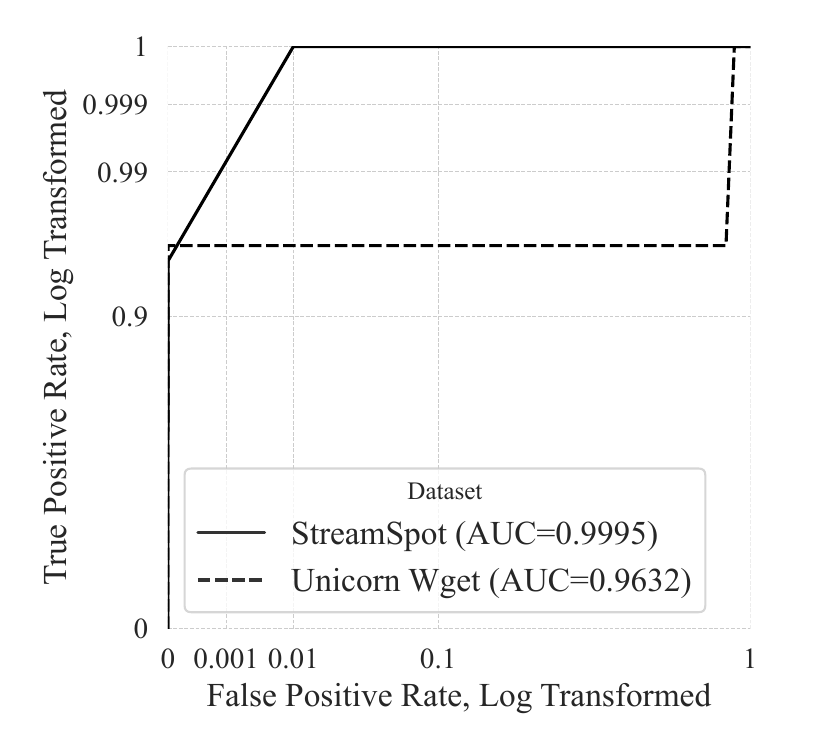}}
    \subfigure[ROC curves on system entity level detection]{\includegraphics[width=0.495\hsize]{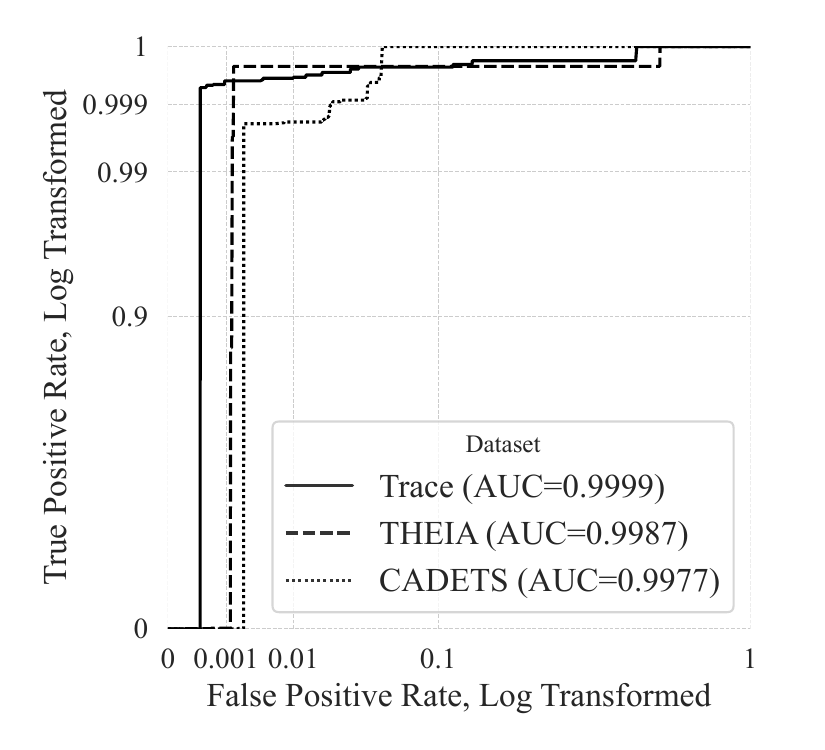}}
    \caption{ROC curves on all datasets.}
    \label{fig_roc_revised}
\end{figure}

\noindent\textbf{Detection Result.} Results show that \system{} successfully detects APTs with high accuracy in different scenarios. We present the detection results of \system{} on each dataset in Table~\ref{tab_general_result_revised} and their corresponding ROC curves in Figure~\ref{fig_roc_revised}.

\begin{table*}[t]
\caption{\system{}'s detection results on different datasets. For batched log level detection, the detection targets are log pieces. And for system entity level detection, system entities are the targets.}
\label{tab_general_result_revised}
\resizebox{\linewidth}{!}{
\begin{tabular}{|c|c|c|c|c|c|c|c|c|c|c|c|c|c|c|}
\hline
\multirow{2}{*}{Granularity} & \multicolumn{2}{c|}{\multirow{2}{*}{Dataset}} & \multirow{2}{*}{Train Ratio} & \multicolumn{2}{c|}{Ground Truth} &\multirow{2}{*}{\#TP} &\multirow{2}{*}{\#FP} & \multirow{2}{*}{\#TN} & \multirow{2}{*}{\#FN} & \multirow{2}{*}{Precision} & \multirow{2}{*}{Recall} & \multirow{2}{*}{FPR} & \multirow{2}{*}{F1-Score} & \multirow{2}{*}{AUC}\\
\cline{5-6}

 & \multicolumn{2}{c|}{} & & \#Benign & \#Malicious & & & & & & & & &\\
\hline\hline
 
 \multirow{2}{*}{Batched log level} & \multicolumn{2}{c|}{StreamSpot} & 80\% & 100 & 100 & 100.0 & 0.59 & 99.41 & 0.0 & 99.41\% & 100.00\% & 0.59\% & 99.71\% & 99.95\%\\
 \cline{2-15}

 & \multicolumn{2}{c|}{Unicorn Wget} & 80\% & 25 & 25 & 24.0 & 0.5 & 24.5 & 1.0 & 98.02\% & 96.00\% & 2.00\% & 96.98\% & 96.32\%\\
 \hline\hline

 \multirow{6}{*}{System entity level} & \multirow{4}{*}{DARPA E3 Trace} & All & \multirow{4}{*}{80\%} & \multirow{4}{*}{616,025} & 68,082 & 68,072 & \multirow{4}{*}{569} & \multirow{4}{*}{615,456} & 10 & \multirow{4}{*}{99.17\%} & \multirow{4}{*}{99.98\%} & \multirow{4}{*}{0.09\%} & \multirow{4}{*}{99.57\%} & \multirow{4}{*}{99.99\%}\\
 \cline{3-3}\cline{6-7}\cline{10-10}

 & & Extension Backdoor & & & 732 & 727 & & & 5 & & & & &\\
 \cline{3-3}\cline{6-7}\cline{10-10}

 & & Pine Backdoor & & & 67,345 & 67,342 & & & 3 & & & & &\\
 \cline{3-3}\cline{6-7}\cline{10-10} 

 & & Phishing Executable & & & 5 & 3 & & & 2 & & & & &\\
 \cline{2-15}

 & DARPA E3 THEIA & All & 80\% & 319,448 & 25,319 & 25,318 & 456 & 318,992 & 1 & 98.23\% & 99.99\% & 0.14\% & 99.11\% & 99.87\%\\
 \cline{2-15}

 & DARPA E3 CADETS & All & 80\% & 344,327 & 12,846 & 12,816 & 759 & 343,568 & 30 & 94.40\% & 99.77\% & 0.22\% & 97.01\% & 99.77\%\\
 \hline

\end{tabular}}
\end{table*}

\begin{figure}[t]
    \centering
    \includegraphics[width=0.50\textwidth]{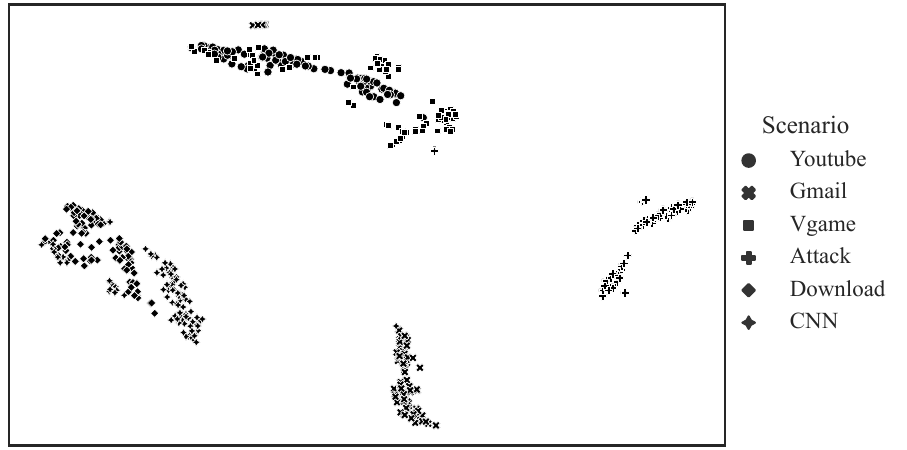}
    \caption{Latent space of system state embeddings in the StreamSpot dataset. Each point represents a log piece in the dataset, which belongs to one of six scenarios: watching YouTube, checking G-mail, playing vgame, undergoing a drive-by-download attack, downloading ordinary files and watching CNN.}
    \label{fig_scatter_revised}
  \end{figure}

On easy datasets such as the StreamSpot dataset, \system{} achieves almost perfect detection results. This is because the StreamSpot dataset collects only single user activity per log batch, resulting in system behaviors that can be easily separated from each other. We further present this effect by visualizing the distribution of system state embeddings abstracted from those log batches in Figure~\ref{fig_scatter_revised}. The system state embeddings are separated into 6 categories, matching the 6 scenarios involved in the dataset. Also, this indicates that \system{}'s graph representation module excels at abstracting system behaviors into such embeddings.

When dealing with the Unicorn Wget dataset, \system{} yields an average 98.01\% precision and 96.00\% recall, significantly lower compared to that of the StreamSpot dataset. \system{}'s self-supervised style makes it difficult to distinguish between stealth attacks and benign system behaviors. However, \system{} still successfully recovers an average 24 out of 25 log batches with only 0.5 false positives generated, better than any of the state-of-the-art detectors~\cite{aptlearnbased-10,aptlearnbasedprovgem-15,aptlearnbasedthreatrace-17}.

On the DARPA datasets, \system{} achieves an average 99.91\% recall and 0.15\% false positive rate with only benign log entries for training. This indicates \system{} quickly learns to model system behaviors. The test set in this scenario is unbalanced, which means the total number of ground-truth benign entities far exceeds that of malicious entities. Among 1,386,046 test entities, only 106,246 are labeled malicious. However, few false positives are generated, as \system{} identifies malicious entities with outlier detection and anomalous entities are naturally detected as abnormalities.

Among the false negative results, we notice that most of them are malicious files and libraries involved in the attacks. This indicates that \system{} excels at detecting malicious processes and network connections that behave differently from benign system entities. However, \system{} finds it hard to locate passive entities such as malicious files and libraries, whose behaviors tend to be similar to benign ones. Fortunately, these intermediate files and libraries can be easily identified during attack story recovery, given the malicious processes and connections successfully detected.

\noindent\textbf{\system{} vs. \emph{State-of-the-art}.} 
The three datasets used to evaluate \system{} are also used by several state-of-the-art learning-based APT detection approaches. For instance, Unicorn~\cite{aptlearnbased-10}, Prov-Gem~\cite{aptlearnbasedprovgem-15} and ThreaTrace~\cite{aptlearnbasedthreatrace-17} for Unicorn Wget and StreamSpot dataset, ThreaTrace and ShadeWatcher~\cite{aptlearnbasedshadewatcher-18} for sub-dataset E3-Trace. Methods that require \emph{a priori} information about APTs, such as Holmes~\cite{aptrulebasedholmes-3}, Poirot~\cite{aptrulebased-5} and Morse~\cite{aptrulebased-6}, are not taken into consideration as \system{} cannot be compared to them in the same detection scenario.

Comparison results between \system{} and other state-of-the-art approaches on each dataset are presented in Table~\ref{tab_compare_result_revised}. Comparison between \system{} and other unsupervised approaches (i.e. Unicorn and ThreaTrace) yields a total victory of our approach, revealing \system{}'s effectiveness in modeling and detecting outliers of benign system behaviors with no supervision from attack-containing logs. 

\begin{table}[t]
\setlength{\abovecaptionskip}{0cm}
\setlength{\belowcaptionskip}{0.2cm}
\caption{Comparison between \system{} and state-of-the-art APT detection methods on different datasets. Within column \emph{supervision}, B indicates benign data, A refers to attack data and SA for streaming attack data.}
\label{tab_compare_result_revised}

\resizebox{\columnwidth}{!}{
\begin{tabular}{|c|c|c|c|c|c|c|c|}

\hline
Dataset & Approach & Train Ratio & Supervision & Precision &  F1-Score & Recall & FPR\\
\hline\hline

\multirow{5}{*}{StreamSpot} & StreamSpot & 80\% & B & 73\% & 81\% & 91\% & 6.6\%\\

 & Unicorn (baseline) & 75\% & B & 95\% & 96\% & 93\% & 1.6\%\\

 & Prov-Gem & 80\% & B,A & 100\% & 97\% & 94\% & \textbf{0\%}\\

 & ThreaTrace & 75\% & B & 98\% & 99\% & 99\% & 0.4\%\\

 & \textbf{\system{} (Ours)} & 80\% & B & 99\% & \textbf{99\%} & \textbf{100\%} & 0.6\%\\
\hline\hline

\multirow{4}{*}{\makecell[c]{Unicorn\\Wget}} & Unicorn (baseline) & 80\% & B & 86\% & 90\% & 95\% & 15.5\%\\

 & Prov-Gem & 80\% & B,A & 100\% & 89\% & 80\% & \textbf{0\%}\\

 & ThreaTrace & 80\% & B & 93\% & 95\% & \textbf{98\%} & 7.4\%\\

 & \textbf{\system{} (Ours)} & 80\% & B & 98\% & \textbf{97\%} & 96\% & 2.0\%\\
\hline\hline

\multirow{5}{*}{\makecell[c]{DARPA\\E3 Trace}} & DeepLog & N/A & B,A & 41\% & 51\% & 68\% & 2.7\%\\

& Log2vec (baseline) & N/A & B,A & 54\% & 64\% & 78\% & 1.8\%\\

& ThreaTrace & N/A & B & 72\% & 83\% & 99\% & 1.1\%\\

& ShadeWatcher & 80\% & B,SA & 97\% & 99\% & \textbf{99\%} & 0.3\%\\

 & \textbf{\system{} (Ours)} & 80\% & B & \textbf{99\%} & \textbf{99\%} & 99\% &\textbf{0.1\%}\\
\hline\hline

\multirow{4}{*}{\makecell[c]{DARPA\\E3 THEIA}} & DeepLog & N/A & B,A & 16\% & 15\% & 14\% & 0.5\%\\

& Log2vec (baseline) & N/A & B,A & 62\% & 64\% & 66\% & 0.3\%\\

& ThreaTrace & N/A & B & 87\% & 93\% & 99\% & \textbf{0.1\%}\\

 & \textbf{\system{} (Ours)} & 80\% & B & \textbf{98\%} & \textbf{99\%} & \textbf{99\%} & 0.1\%\\
\hline\hline

\multirow{4}{*}{\makecell[c]{DARPA\\E3 CADETS}} & DeepLog & N/A & B,A & 23\% & 35\% & 74\% & 4.4\%\\

& Log2vec (baseline) & N/A & B,A & 49\% & 62\% & 85\% & 1.6\%\\

& ThreaTrace & N/A & B & 90\% & 95\% & \textbf{99\%} & 0.2\%\\

 & \textbf{\system{} (Ours)} & 80\% & B & \textbf{94\%} & \textbf{97\%} & 99\% & \textbf{0.2\%}\\
\hline
\end{tabular}
}
\end{table}

Beyond unsupervised methods, Prov-Gem is a supervised APT detector based on GATs. However, it fails to achieve a better detection result on even the easiest StreamSpot dataset. This is mainly because simple GAT layers supervised on classification tasks are not as expressive as graph masked auto-encoders in producing high-quality embeddings. Another APT detector mentioned, ShadeWatcher, adopts a semi-supervised detection approach. ShadeWatcher leverages TransR~\cite{transr-43} and graph neural networks (GNNs) to detect APTs based on recommendation and is able to achieve the best recall rate on the E3-Trace sub-dataset. TransR, a self-supervised graph representation method on Knowledge Graphs, contributes most to the detection accuracy of ShadeWatcher. Unfortunately, TransR is extremely expensive in computation overhead. For example, ShadeWatcher spends as much as 12 hours training the TransR module on the E3-Trace sub-dataset. On the contrary, evaluation on the computation overhead of \system{}(Sec.~\ref{sec_evaluation_4_revised}) shows that \system{} is able to complete training on the same amount of training data 51 times faster than ShadeWatcher~\cite{aptlearnbasedshadewatcher-18}.

\subsection{False Positive Analysis}\label{sec_evaluation_3_revised}

For real-time applications, APT detectors must prevent false alarms at best effort, as those false alarms often tire and confuse security analysts. We evaluate \system{}'s false positive rate (FPR) on benign audit logs and investigate how our model adaption mechanism reduces those false alarms. Table~\ref{tab_general_result_revised} shows \system{}'s false positive rate on each dataset. Within each dataset, only benign logs are used for training and testing. \system{} yields low FPR (average 0.15\%) with large training data. This is because \system{} models benign system behaviors with self-supervised embeddings, allowing it to effectively handle unseen system entities. Such a low FPR enables \system{}'s application under real-world settings. When conducting fine-grain entity-level detection only, \system{} only yields 569 false alarms on the Trace dataset, with an average only 40 false alarms every day. Security analysts can easily handle this number of alarms and do security investigations on them. If the two-stage detection described in Sec.~\ref{sec_overview_revised} is applied, the average number of false alarms every day can be further lowered to 24.

Our model adaption mechanism is designed to help \system{} to learn from newly-observed unseen behaviors. We evaluate how such mechanism reduces false positives on benign audit logs in Table~\ref{tab_fp_result_revised}. Specifically, we test \system{} on the Trace dataset under five different settings:

\vspace{3pt}\noindent$\bullet$~Training on the first 80\% log entries and testing on the rest 20\% with no adaption, identical to our original setting.

\vspace{3pt}\noindent$\bullet$~Training on the first 20\% and testing on the last 20\% with no adaption, for comparison purpose.

\vspace{3pt}\noindent$\bullet$~Training on the first 20\%, adapting on false positives generated from the following 20\%, and testing on the last 20\%.

\vspace{3pt}\noindent$\bullet$~Training on the first 20\%, adapting on both false positives and true negatives generated from the following 20\% log entries, and testing on the last 20\%.

\vspace{3pt}\noindent$\bullet$~Training on the first 20\%, adapting on both false positives and true negatives generated from the following 40\% log entries, and testing on the last 20\%.

\begin{table}[t]
\setlength{\abovecaptionskip}{0cm}
\setlength{\belowcaptionskip}{0.2cm}
\caption{\system{}'s false positive rates on different datasets. The effect of model adaption mechanism is tested under different settings.}
\label{tab_fp_result_revised}
\centering
\resizebox{0.8\columnwidth}{!}{
\begin{tabular}{|c|c|c|c|c|}

\hline
Dataset & Train Ratio & Adaption & Test Ratio & FPR\\
\hline\hline

StreamSpot & 80\% & N/A & 20\% & 0.59\%\\
\hline

Unicorn Wget & 80\% & N/A & 20\% & 2.00\%\\
\hline

\multirow{5}{*}{\makecell[c]{DARPA\\E3 Trace}} & 80\% & N/A & 20\% & 0.089\%\\

& 20\% & N/A & 20\% & 0.426\%\\

& 20\% & 20\% FP & 20\% & 0.272\%\\

& 20\% & 20\% FP \& TN & 20\% & 0.220\%\\

& 20\% & 40\% FP \& TN & 20\% & 0.173\%\\
\hline
\end{tabular}}
\end{table}

Experimental results indicate that adapting the model to benign feedbacks consistently reduces false positives. A further reduction can be achieved by feeding both false positives and true negatives to the model. This is because the graph representation module can be retrained with any data to enhance its representation ability, as described in Sec~\ref{sec_model_4_revised}.

\subsection{Performance Overhead}\label{sec_evaluation_4_revised}

\system{} is designed to perform APT detection with minimum overhead, granting it applicability under various conditions. \system{} completes its training and inference in logarithmic time and takes up linear space. We provide a detailed analysis of its time and space complexity in Appendix~\ref{appe_b_revised}. We further test \system{}'s run-time performance on sub-dataset E3-Trace and present its time and memory consumption in Table~\ref{tab_overhead_revised}.

\begin{table}[t]
\centering
\setlength{\belowcaptionskip}{0.2cm}
\caption{Performance overhead of \system{} on the E3-Trace sub-dataset.}
\label{tab_overhead_revised}
\resizebox{\columnwidth}{!}{
\begin{tabular}{|c|c|c|c|c|c|}
\hline
\multirow{2}{*}{Phase} & \multirow{2}{*}{Component} & \multicolumn{2}{c|}{Time consumption (s)} & \multirow{2}{*}{Peak Memory consumption (MB)}\\
\cline{3-4}

& & with GPU & CPU only & \\

\hline\hline
\makecell[c]{Graph\\Construction} & N/A & \multicolumn{2}{c|}{642}  & 2,610\\
\hline

\multirow{2}{*}{Training} & \makecell[c]{Graph\\Representation} & 151 & 685 & 1,564\\

\cline{2-5}

& Detection & \multicolumn{2}{c|}{78} & 1,320\\
\hline

\multirow{2}{*}{Inference} & \makecell[c]{Graph\\Representation} & 5 & 10 & 2,108\\

\cline{2-5}

& Detection & \multicolumn{2}{c|}{825} & 1,667\\
\hline
\end{tabular}}
\end{table}
In real-world settings, GPUs may not be available at all. We also test \system{}'s efficiency without the GPU. With only CPUs available, the training phase becomes apparently slower. The efficiency of graph construction and outlier detection phase is not affected as they are implemented to perform on CPUs. We also measure the max memory consumption during training and inference. \system{}'s low memory consumption not only prevents OOM problems on huge datasets, but also makes \system{} approachable under tight conditions.

Evaluation results on performance overhead manifest the claim that \system{} is advantageous over other state-of-the-art APT detectors in efficiency. For instance, ATLAS~\cite{aptlearnbasedatlas-11} takes about an hour to train its model on 676MB of audit logs and ShadeWatcher~\cite{aptlearnbasedshadewatcher-18} takes 1 day to train on the E3-Trace sub-dataset. Compared with ShadeWatcher, \system{} is 51 times faster in training under the same train ratio setting (i.e. 80\% log entries for training on E3-Trace sub-dataset).

These evaluation results also illustrate that \system{} is fully practicable in different conditions. Considering the fact that sub-dataset E3-Trace is collected in 2 weeks, 1.37GB of audit logs is produced per day. This means under CPU-only conditions, \system{} takes only 2 minutes to detect APTs from those logs and complete model adaption every day. Such promising efficiency makes \system{} an available choice for individuals and small-size enterprises. For larger enterprises and institutions, they produces audit logs in hundreds of GBs every day~\cite{datascale-23}. In this case, efficiency of \system{} can be ensured by training and adapting itself with GPUs and parallelizing the detection module with distributed CPU cores.

\subsection{Ablation Study}\label{sec_evaluation_5_revised}

In this section, we first address the effectiveness of important individual components in \system{}'s graph representation module, then carry out a hyper-parameter analysis to evaluate the sensitivity of \system{}. Analysis on individual components and most hyper-parameters are conducted on the most difficult Wget dataset and the hyper-parameter analysis on the detection threshold $\theta$ is conducted on all datasets.

\begin{figure}[t]
    \centering
    \subfigure[FR, on Wget dataset]{\includegraphics[width=0.495\hsize]{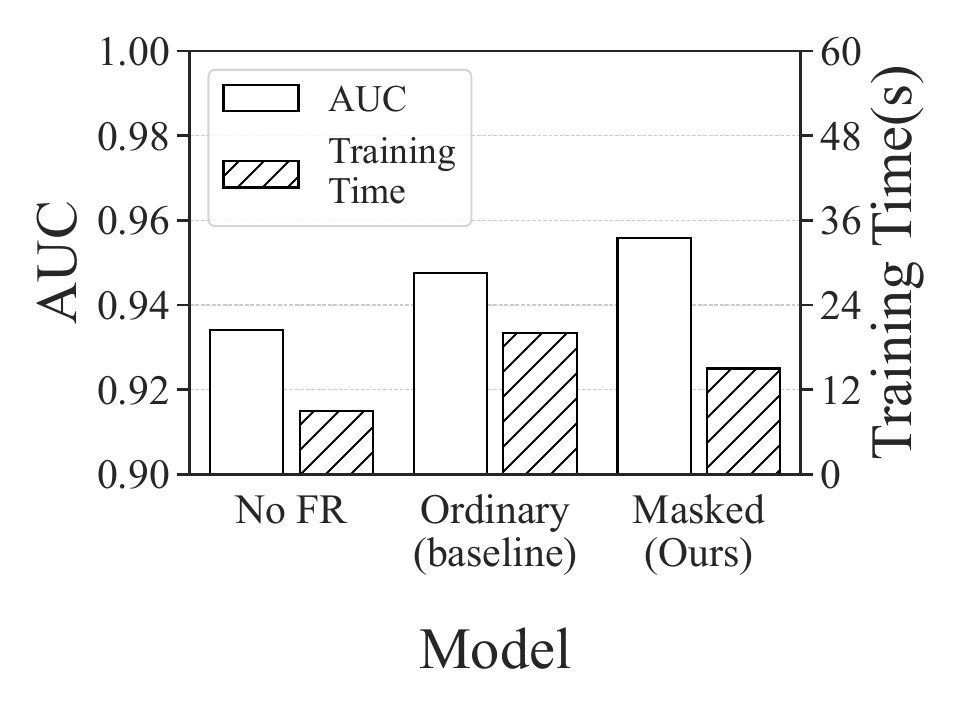}\label{fig_ablation_fr}}
    \subfigure[SR, on Wget dataset]{\includegraphics[width=0.495\hsize]{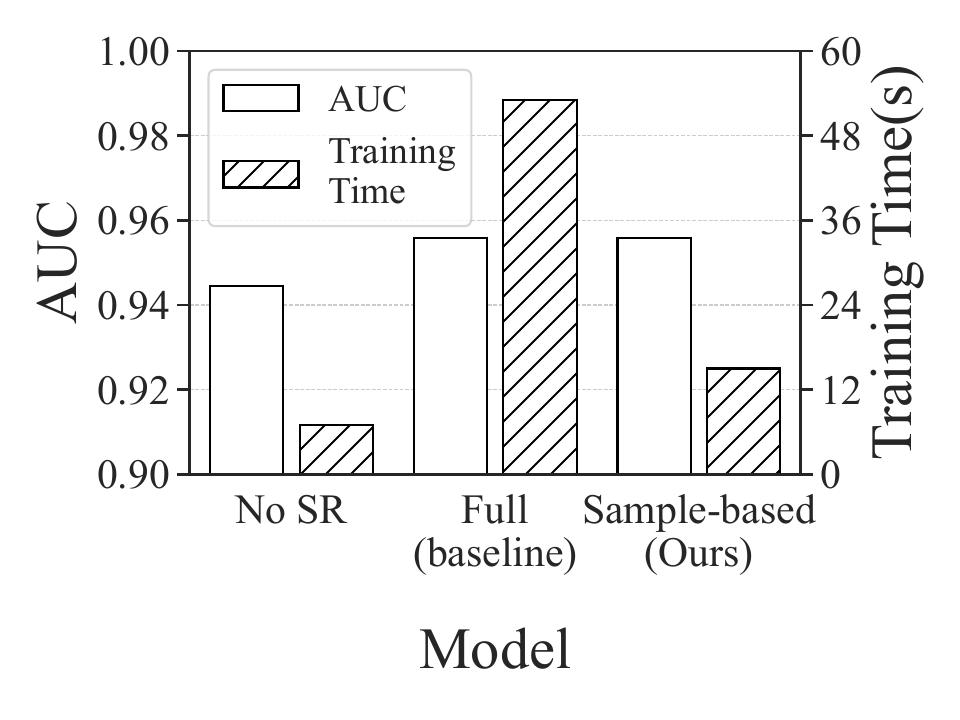}\label{fig_ablation_sr}}
    \caption{Effect of different reconstruction components on \system{}'s performance and efficiency.}
    \label{fig_ablation_revised}
\end{figure}

\noindent\textbf{Individual Component Analysis.} We study how feature reconstruction (FR) as well as structure reconstruction (SR) affect \system{}'s performance and Figure~\ref{fig_ablation_revised} presents the impact of these component to both detection result and performance overhead. Both FR and SR provide supervision for \system{}'s graph representation module. Compared with ordinary FR, Masked FR slightly boosts performance and significantly reduces training time. Sampled-based SR, however, is an effective complexity reduction component which accelerates training without losing performance, compared with full SR.

\begin{figure*}[t]
    \centering
    \setcounter{subfigure}{0}
    \subfigure[\textbf{\emph{d}}, on Wget dataset]{\includegraphics[width=0.3\hsize]{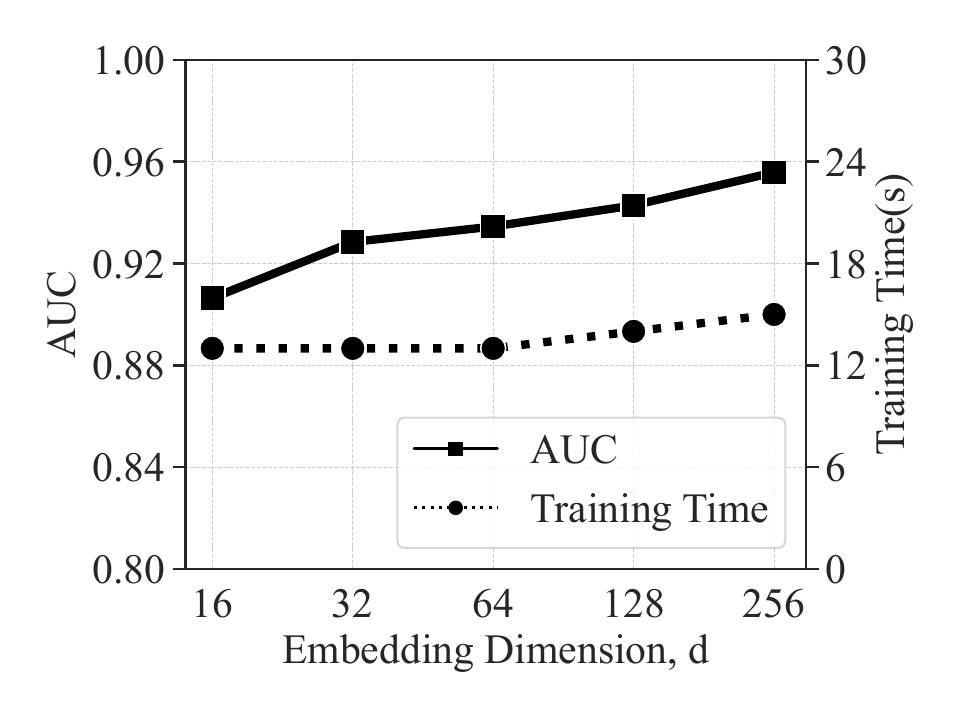}\label{fig_d_revised}}
    \subfigure[\textbf{\emph{l}}, on Wget dataset]{\includegraphics[width=0.3\hsize]{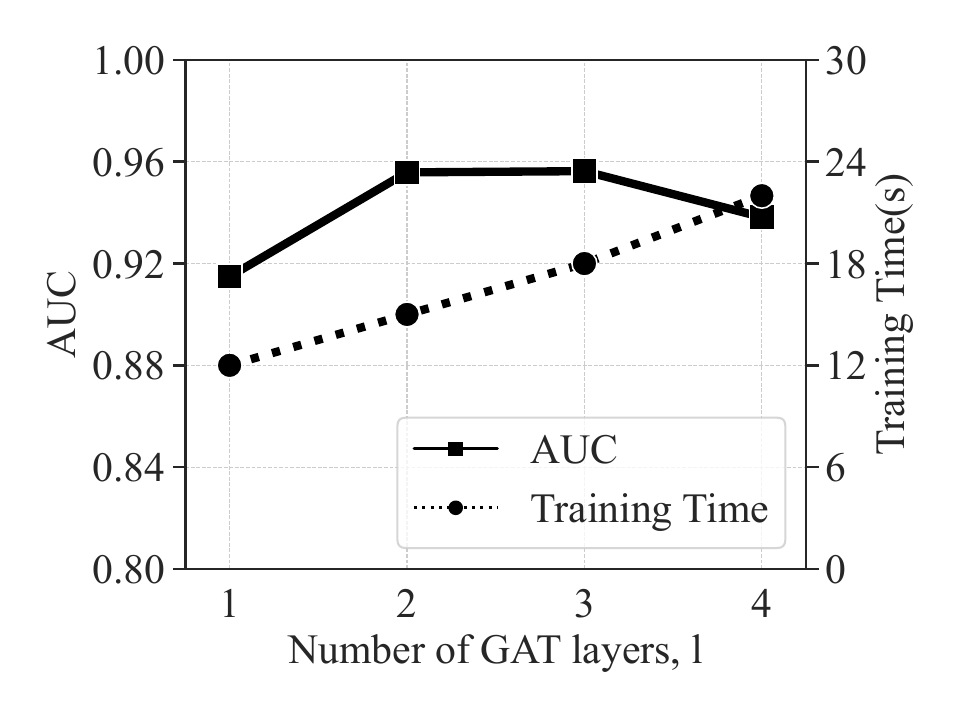}\label{fig_l_revised}}
    \subfigure[\textbf{\emph{r}}, on Wget dataset]{\includegraphics[width=0.3\hsize]{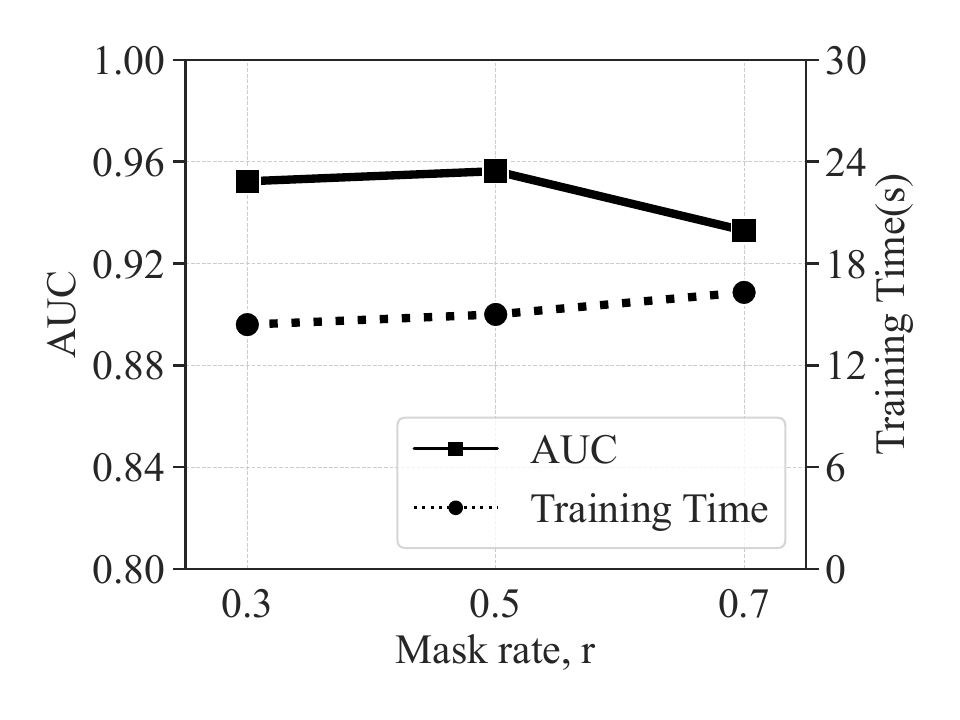}\label{fig_r_revised}}
    \caption{Effect of different hyper-parameters on \system{}'s performance and efficiency.}
    \label{fig_hyperparameter}
\end{figure*}

\noindent\textbf{Hyper-parameter Analysis.} The function of \system{} is controlled by several hyper-parameters, including an embedding dimension \emph{d}, number of GAT layers \emph{l}, node mask rate \emph{r} and the outlier detection threshold $\theta$. Figure~\ref{fig_hyperparameter} illustrates how these hyper-parameters affect model performance in different situations. Hyper-parameters have little impact in most cases. 

Generally speaking, relatively higher model performance is achieved with a larger embedding dimension and more GAT layers by collecting more information from more distant neighborhoods, as shown in Sub-figure~\ref{fig_d_revised} and~\ref{fig_l_revised}. However, increasing $d$ or $l$ introduces heavier computation which leads to longer training and inference time.

The default mask rate of $0.5$ yields best results. This is because \system{} is unable to get sufficient training under a low mask rate. And under a high mask rate, node features are severely damaged which prevents \system{} to learn node embeddings via feature reconstruction. Increasing mask rate slightly introduces more computation burden.

\begin{figure}[t]
    \centering
    \includegraphics[width=0.3\textwidth]{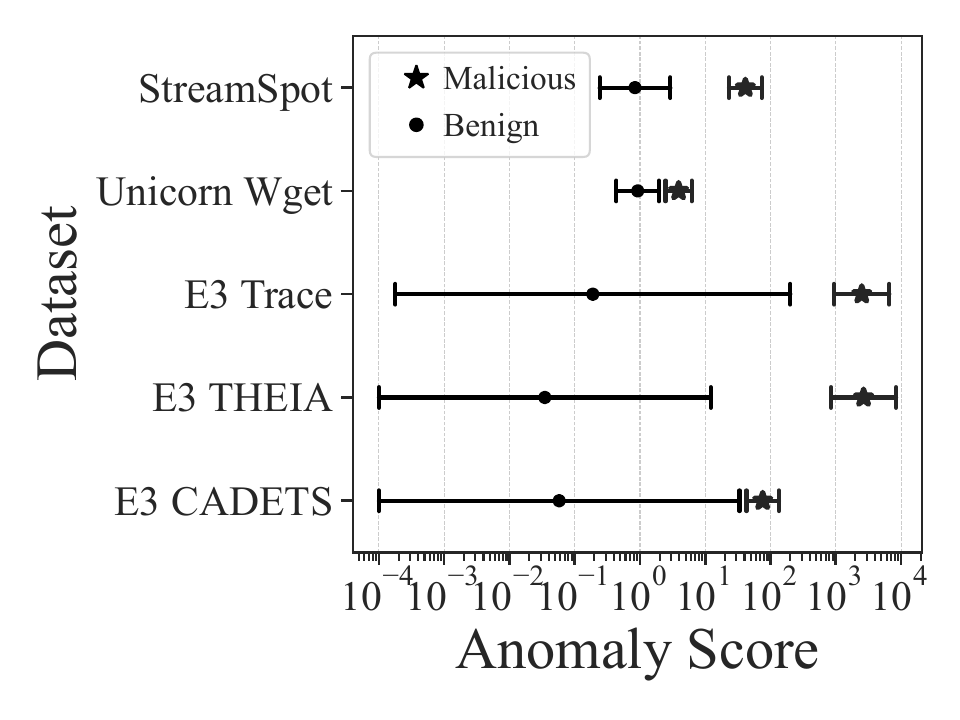}
    \caption{Anomaly scores of system entities. We exclude the highest and lowest 5\% scores on each dataset.}
    \label{fig_score_revised}
\end{figure}

We further examine the anomaly scores of entities to assess the sensitive of $\theta$. A lower $\theta$ naturally leads to a higher recall performance at the cost of more false positives and vice versa. As demonstrated in Figure~\ref{fig_score_revised}, most malicious entities are given high anomaly scores compared with benign ones and are well-separated from them with little overlapping. The considerable spaces between benign and malicious anomaly scores support the claim that \system{} does not depend on a precise threshold $\theta$ to perform accurate detection in practical situations. We quantify such spaces in Appendix~\ref{appe_f_revised}.

For an unsupervised detector as \system{}, hyper-parameters are usually difficult to select. However, that is not the case for \system{}. We provide a general guideline for hyper-parameter selection also in Appendix~\ref{appe_f_revised}.

\section{Discussion and Limitations}\label{sec_discussion_revised}

\noindent\textbf{Quality of Training Data.} \system{} models benign system behaviors and detects APTs with outlier detection. Similar to other anomaly-based APT detection approaches~\cite{aptlearnbased-10,aptstatbasednodoze-7, aptstatbased-8}, we assume that all up-to-date benign system behaviors are observed during training log collection. However, if \system{} is trained with low quality data that insufficiently covers system behaviors, many false positives can be generated. .

\noindent\textbf{Outlier Detection.} \system{} implements a KNN-based outlier detection module for APT detection. While it completes training and inference in logarithmic time, its efficiency on large datasets is still unsatisfactory. To illustrate, our detection module takes 13.8 minutes to check 684,111 targets, making up 99\% of total inference time. Other outlier detection methods, such as One-class SVM~\cite{ocsvm-47} and Isolation Forest~\cite{if-48}, do not fit in our detection setting and cannot adapt to concept drift. Cluster-based methods and approximate KNN search may be more suitable to huge datasets. Meanwhile, KNN-based methods may be extended to GPU. We leave such improvements on our detection module to future research.

\noindent\textbf{Adversarial Attacks.} In Sec.~\ref{sec_evaluation_2_revised}, we show that \system{} handles stealth attacks well, which avoid detection by behaving similar to the benign system. However, if attackers get to know how \system{} works in details, they might conduct elaborately designed attacks to infiltrate our detector. We make a simple analysis on adversarial attacks and demonstrate \system{}'s robustness against them in Appendix~\ref{appe_a_revised}. As Graph-based approaches become increasingly popular and powerful in different detection applications, designing and avoiding these adversarial attacks on both the input graphs and the GNNs stands for an interesting research topic.

\section{Related Work}

\system{} is mainly related to three fields of research, including the detection of APTs, graph representation learning and outlier detection methods.

\noindent\textbf{APT Detection.} The goal of APT detection is to detect APT signals, malicious entities and invalid interactions from audit logs. Recent works are mostly based on data provenance. As suggested by~\cite{aptlearnbasedshadewatcher-18}, provenance-based detectors can be categorized into rule-based, statistics-based and learning-based approaches. Rule-based approaches~\cite{aptrulebased-2,aptrulebasedholmes-3,aptrulebased-4,aptrulebased-5,aptrulebased-6} utilize \emph{a priori} knowledge about previous seen attacks and construct unique heuristic rules to detect them. Statistics-based approaches~\cite{aptstatbased-8,aptstatbased-9,aptstatbasednodoze-7} construct statistics to measure the abnormality of provenance graph elements and perform anomaly detection on them. Learning-based approaches~\cite{aptlearnbased-10,aptlearnbasedatlas-11,aptlearnbased-12,atplearnbased-13,aptlearnbased-14,aptlearnbasedprovgem-15,aptlearnbased-16,aptlearnbasedthreatrace-17,aptlearnbasedshadewatcher-18,aptlearnbasedsigl-58} leverage deep learning to model either benign system behaviors~\cite{aptlearnbased-10,aptlearnbased-12,aptlearnbasedthreatrace-17} or attack patterns~\cite{aptlearnbasedatlas-11,aptlearnbased-14,aptlearnbasedprovgem-15,aptlearnbased-16,aptlearnbasedshadewatcher-18} and perform APT detection as classification~\cite{aptlearnbasedatlas-11,aptlearnbasedprovgem-15,aptlearnbased-16} or anomaly detection~\cite{aptlearnbasedshadewatcher-18,aptlearnbasedsigl-58}. Among them, sequence-based methods~\cite{aptlearnbasedatlas-11,aptlearnbased-14} detect APTs based on system execution/workflow patterns and graph-based methods~\cite{aptlearnbased-10,aptlearnbased-12,aptlearnbasedprovgem-15,aptlearnbased-16,aptlearnbasedthreatrace-17,aptlearnbasedshadewatcher-18,aptlearnbasedsigl-58} model entities and interactions via GNNs and detect abnormal behaviors as APTs.

\noindent\textbf{Graph Representation Learning.} Embedding techniques on graphs start from the graph convolutional network (GCN)~\cite{gcn-51} and are further enhanced by the graph attention network (GAT)~\cite{gat-29} and GraphSAGE~\cite{graphsage-52}. Graph auto-encoders~\cite{vgae-24,gala-25,gate-26,gmae-19} bring unsupervised solutions for graph representation learning. GAE, VGAE~\cite{vgae-24} and GATE~\cite{gate-26} utilize feature reconstruction and structure reconstruction to learn output embeddings in a self-supervised way. However, they focus on link prediction and graph clustering tasks, irrelevant to our application. Recently, graph masked auto-encoders~\cite{gmae-19} leverage masked feature reconstruction and have achieved state-of-the-art performance on various applications.

\noindent\textbf{Outlier Detection.} Outlier detection methods view outliers (i.e. objects that may not belong to the ordinary distribution) as anomalies and aim to identify them. Traditional outlier detection methods include One-class SVMs~\cite{ocsvm-47}, Isolation Forest~\cite{if-48} and Local Outlier Factor~\cite{lof-53}. These traditional approaches are widely used in various detection scenarios, including credit card fraud detection~\cite{ocan-55} and malicious transaction detection~\cite{transaction-57}. This proves that outlier detection methods work well in anomaly detection. Meanwhile, auto-encoders themselves are effective tools for outlier detection. We explain why we do not use graph auto-encoders as anomaly detectors in Appendix~\ref{appe_d_revised}.
\section{Conclusion}\label{sec_conclusion}

We have introduced \system{}, an universally applicable APT detection approach that operates in utmost efficiency with little overhead. \system{} leverages masked graph representation learning to model benign system behaviors from raw audit logs and performs multi-granularity APT detection via outlier detection methods. Evaluations on three widely-used datasets under various detection scenarios indicate that \system{} achieves promising detection results with low false positive rate and minimum computation overhead.
\newpage
\section*{Acknowledgments}

We would like to thank the anonymous reviewers and our shepherd for their detailed and valuable comments. This work was supported by the National Natural Science Foundation of China (No.~U1936213 and No.~62032025), CNKLSTISS, the Fundamental Research Funds for the Central Universities, Sun Yat-sen University (No.~22lgqb26) and Program of Shanghai Academic Research Leader (No.~21XD1421500).

\bibliographystyle{unsrt}
\bibliography{ref}

\section*{Appendix}
\appendix

\section{Analysis on Adversarial Attack}\label{appe_a_revised}
Adversarial attacks against \system{}, such as evasion attacks and poison attacks, are tricky to implement but still possible to carry out. Two types of adversarial attacks are potentially practical against \system{}, manipulating input audit logs or exploiting model architecture. The later approach is not an option for \system{}'s attackers, as they have no access to \system{}'s inner parameters. Similar to SIGL~\cite{aptlearnbasedsigl-58}, we conduct a simple experiment concerning \system{}'s robustness against graph manipulations, which shows these attacks do not affect \system{}'s detection effectiveness. 

In this experiment, we expect attackers have no knowledge of \system{}'s inner parameters and cannot get any feedback from \system{}. However, attackers can freely manipulate the malicious entities within \system{}'s input audit logs, as well as a small proportion of benign entities. Consequently, we consider four types of attacks:

\textbf{Malicious Feature Evasion (MFE).} Attackers have altered the features of all malicious entities in the raw audit logs trying to evade detection. This affects the node initial embeddings of the input provenance graph and forces malicious entities to mimic benign ones in node features.

\textbf{Malicious Structure Evasion (MSE).} Attackers have adding new edges between malicious entities and benign ones, so that malicious entities behave more normally and tend to have local structures more similar to benign entities. This affects the graph representation module and pulls the embeddings of malicious nodes towards benign ones, making them more difficult to identify.

\textbf{Combined Evasion (MCE).} This type of attack is a combination of the MFE and MSE and causes malicious entities to approximate benign ones in both features and structures.

\textbf{Benign Feature Poison (BFP).} Attackers have manipulated the features of benign entities to poison \system{}. Attackers have injected some benign entities with similar initial features as malicious entities, trying to convince \system{} that the malicious entities behave normally. This shifts benign entities to mimic malicious ones in node features and gradually poisons \system{}'s detection module.

We evaluate \system{}'s robustness against the above four attack strategies on the E3-Trace sub-dataset and we present the results in Table~\ref{tab_adversarial}. Experimental results confirm that \system{} is robust enough against adversarial attacks. MFE has almost no impact on \system{}'s effectiveness. \system{} is slightly more vulnerable facing the structure evasion attacks, because the graph structure is critical to unsupervised graph representation learning. However, the structure attacks are still weak against \system{}'s detection effectiveness. We believe that the feature and structure reconstruction involved in our graph representation module contributes to this robustness, as it learns a model to reconstruct both node features and its neighboring structure with the information from its neighbors. Consequently, adversarial attacks involving feature and structure altering of malicious entities will fail, and with \system{}'s model itself well-protected, attackers have to either extend their effort on searching and manipulating benign entities, or find other attack strategies against \system{}. Meanwhile, BFP poses a relatively greater threat to \system{}'s performance. However, under BFP, only 0.057 AUC reduction is achieved by manipulating the input feature of an enormous 161,029 benign entities. This level of intervention in the benign system is beyond the capability of ordinary attackers and will definitely leave observable trace. Therefore, an effective BFP attack is also very difficult to carry out.

\begin{table}[t]
\centering
\small
\setlength{\belowcaptionskip}{0.2cm}
\caption{Impact of different adversarial attack strategies on \system{}'s detection effectiveness.}
\label{tab_adversarial}
\begin{tabular}{|c|c|c|c|c|c|}
\hline
Attack Type & None & MFE & MSE & MCE & BFP \\
\hline

AUC & 0.9999 & 0.9999 & 0.9999 & 0.9994 & 0.9942\\
\hline
\end{tabular}
\end{table}
\section{Time and Space Complexity of \system{}}\label{appe_b_revised}
Given number of system entities $N$, number of system interactions $E$, number of possible node/edge labels \emph{t}, graph representation dimension $d$, number of GAT layers $l$ and mask rate $r$ the graph construction steps builds a featured provenance graph in $O((N+E) * t)$ time, masked feature reconstruction is completed in $O((N+E) * d^2 * l * r)$ time and sample-based structure reconstruction takes only $O(N * d)$ time. Training of the detection module takes $O(N *logN * d)$ time to build a K-D Tree and memorize benign embeddings. Detection result of a single target is obtained in $O(logN * d * k)$ time. Thus, the overall time complexity of \system{} during training and inference is $O(N*logN * d * k+E * d ^2 * l * r + (N + E) * t)$.

\system{}'s memory consumption largely depends on $d$ and $t$. The graph representation module takes up $O((N+E)*t)$ space to store a provenance graph and $O((N+E)*(t +d))$ space to generate output embeddings. The detection module takes $O(N*d)$ space to memorize benign embeddings. The overall space complexity of \system{} is $O((N+E)*(t+d))$.

\section{Case Study}\label{sec_evaluation_6_revised}
\begin{figure}[t]
    \centering
    \includegraphics[width=0.46\textwidth]{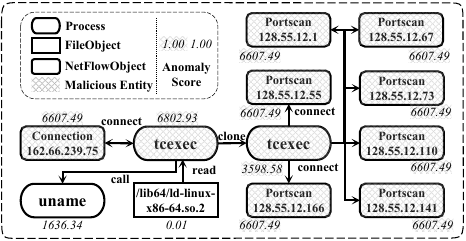}
    \caption{Another provenance graph of our motivating example (i.e. Pine Backdoor). Numbers assigned to nodes are the anomaly scores assessed by \system{}'s detection module.}
    \label{fig_case_revised}
\end{figure}

We use our motivating example described in Sec.~\ref{sec_motivation_1_revised} again to illustrate how \system{} detects APTs from audit logs. Our motivating example involves an APT attack: Pine Backdoor, which implants an malicious executable to a host via phishing e-mail, aiming to perform internal reconnaissance and build a silent connection between the host and the attacker. We perform \emph{system entity level detection} on it and obtain real detection results from \system{}. First, \system{} constructs a provenance graph from raw audit logs. We provide the example provenance graph in Figure~\ref{fig_case_revised} to illustrate. Among them, \emph{tcexec}, \emph{Connection-162.66.239.75} and the portscan \emph{NetFlowObjects} which \emph{tcexec} connects to are malicious entities while the others are benign ones. The graph representation module then obtains their embeddings via the graph masked auto-encoder. The embedding of \emph{tcexec} is calculated by propagating and aggregating information from its multi-hop neighborhood to model its interaction behavior with other system entities. For instance, its 2-hop neighborhood is namely \emph{Connection-162.66.239.75}, \emph{tcexec}'s sub-process, \emph{uname}, \emph{ld-linux-x86-64.so.2} and other portscan \emph{NetFlowObject}s. The detection module subsequently calculates distances between those embeddings and their k-nearest benign neighbors in the latent space and assigns anomaly scores to them. The malicious entities are given very high anomaly scores (3598.58\textasciitilde6802.93) that far exceed our detection threshold (3000) while others present low abnormality (0.01\textasciitilde1636.34). Thus, \system{} successfully detects malicious system entities with no false alarm generated.

\section{Hyper-parameter Choice Guideline}\label{appe_f_revised}
The choice of detection threshold $\theta$ depends on the corresponding dataset \system{} is operating on. However, finding a precise $\theta$ is not necessary on each individual dataset, according to Sec.~\ref{sec_evaluation_5_revised}. Even if there is only benign data available, users may adjust and choose the optimal $\theta$ according to the resulting false positive rate (i.e. use the first $\theta$ when false positive rate is below the desired value). Here we provide a simple analysis on how wide the threshold selection range is and what consequence selecting different $\theta$ may lead to.

\noindent\textbf{Impact of Different $\theta$ Choices.} Altering $\theta$ does not lead to drastic changes in detection results, especially for \#FN. For example, on the E3-Trace sub-dataset, changing $\theta$ from 1000 to 5000 results in 2 more FNs and an 1\% decrease in FPR. On E3-THEIA, altering $\theta$ from 100 to 500 provides 0.5\% FPR decrease and does not change \#FN. On E3-Cadets, setting $\theta$ as 100 instead of 10 leads to a 10\% decreased FPR with only 30 new FNs generated.

\noindent\textbf{Quantifing the Threshold Selection Range.} By limiting FPR below 1\% on entity level tasks, we get a minimum $\theta$ equals 1980 on E3-Trace, 180 on E3-THEIA dataset and 95 on E3-CADETS dataset. The resulting recalls are respectively 0.99985, 0.99996 and 0.99782. Meanwhile, the maximum $\theta$s that ensure recall > 99\% are actually 6600, 1020 and 120 on the three different sub-datasets. The space between the minimum and maximum $\theta$ is big enough and the curves are flat. Consequently, MAGIC does not need a precise $\theta$ to obtain the desirable result and selecting threshold $\theta$ based on false positive rate is practical.

\section{Noise Reduction}\label{appe_c_revised}
\emph{Noise reduction} is widely adopted by various recent works~\cite{aptlearnbasedatlas-11,aptlearnbasedshadewatcher-18,aptlearnbasedthreatrace-17,aptlearnbasedprovgem-15} to reduce the complexity of provenance graphs and remove redundant and useless information. \system{} applies a mild noise reduction approach as \system{} is less sensitive to the scale of the provenance graph and more information is preserved in this way. Compared with noise reduction done in recent works~\cite{aptlearnbasedatlas-11,aptlearnbasedshadewatcher-18}, we neither delete irrelevant nodes nor merge nodes with the same interaction behavior. This is because (1) attack-irrelevant nodes provide information for benign system behaviors and (2) multiple nodes with the same interaction behavior duplicate the information propagated to near-by nodes and impact on their embeddings.

\section{Auto-encoder Based Anomaly Detection}\label{appe_d_revised}
Among traditional applications of machine learning, anomaly detection via auto-encoders is common practice. Typically, auto-encoders are trained to reconstructing a target and minimize its reconstruction loss. Thus, the reconstruction loss of a newly-arrived sample indicates how similar it behaves to training samples and samples with high reconstruction errors are detected as outliers. However, we do not apply auto-encoder-based outlier detection because of two reasons: (1) our sample-based structure reconstruction produces high-variance reconstruction loss on single sample, which prevents stable threshold-based outlier detection and (2) \system{} performs batched log level detection by detecting outliers in system state embeddings, which do not have a reconstruction target and cannot be compared in reconstruction error.

\section{Entity-level Data Labeling on DARPA TC datasets}
\noindent\textbf{A Feasible Labeling Methodology.} Mining and labeling attack-relevant entities in DARPA TC datasets can be extremely effort-consuming, given the fact that the ground truth document they provide is practically unreadable. Recently, Watson~\cite{datalabelingwatson-59} have carrited out a successful attempt to label the E3-Trace sub-dataset. We are able to repeat this labeling methodology on sub-datasets E3-Trace, E3-THEIA and E3-CADETS. The following is a detailed description on this labeling process:

\vspace{3pt}\noindent$\bullet$~Traverse all log entries in the dataset. Among those records, we extract \textbf{process}, \textbf{file} and \textbf{netflow} entities.

\vspace{3pt}\noindent$\bullet$~Extract entity names. Entities' semantic names are stored in different fields. For instance, \emph{Subject.properties.map.name} stores \textbf{process} names in sub-dataset E3-Trace.

\vspace{3pt}\noindent$\bullet$~Mine key attack-relevant entities from the ground truth document. Some attacks were not well-recorded but at least one of the attacks using the same strategy is well-recorded.

\vspace{3pt}\noindent$\bullet$~Match the names of key attack entities with all extracted entities. The matching entities are labeled as positive. Explore the neighborhood of these entities and search for other entities that are involved in the attack. Newly-identified ones are also treated as positives.

\noindent\textbf{The Resulting Ground Truth.} We perform such labeling steps on sub-datasets E3-Trace, E3-THEIA and E3-CADETS and obtain the following ground truth, in the form of descriptive text and attack graphs.

\begin{figure}[t]
    \centering
    \includegraphics[width=0.48\textwidth]{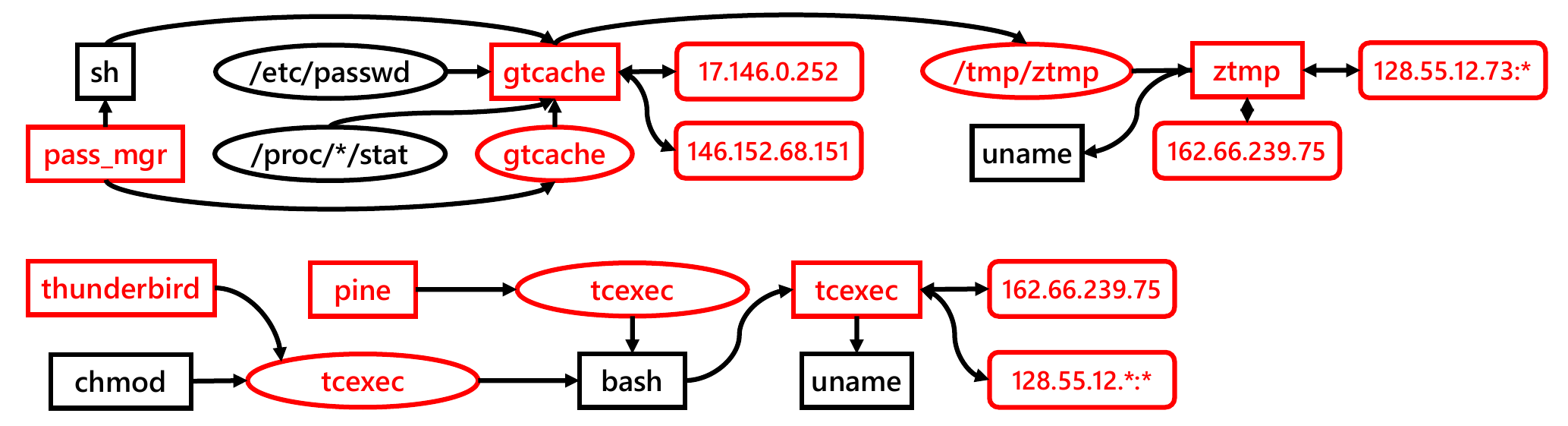}
    \caption{Attack graph of sub-dataset E3-Trace.}
    \label{fig_trace_attack}
\end{figure}
\vspace{3pt}\noindent$\bullet$~E3-Trace (Figure~\ref{fig_trace_attack}). Two successful attack attempts worked on Trace: Browser Extension and Pine Backdoor. During the Browser Extension attack, the user downloaded and executed \emph{gtcache} via browser extension \emph{pass\_mgr}. \textbf{gtcache} communicated with the attacker, scaned sensitive information and created \textbf{ztmp} to portscan host \emph{128.55.12.73}. In the Pine Backdoor attack, the user unfortunately launched the phishing executable \emph{tcexec}. \textbf{tcexec} connected back to the attacker and performed a wide postscan on the local network.

\begin{figure}[t]
    \centering
    \includegraphics[width=0.48\textwidth]{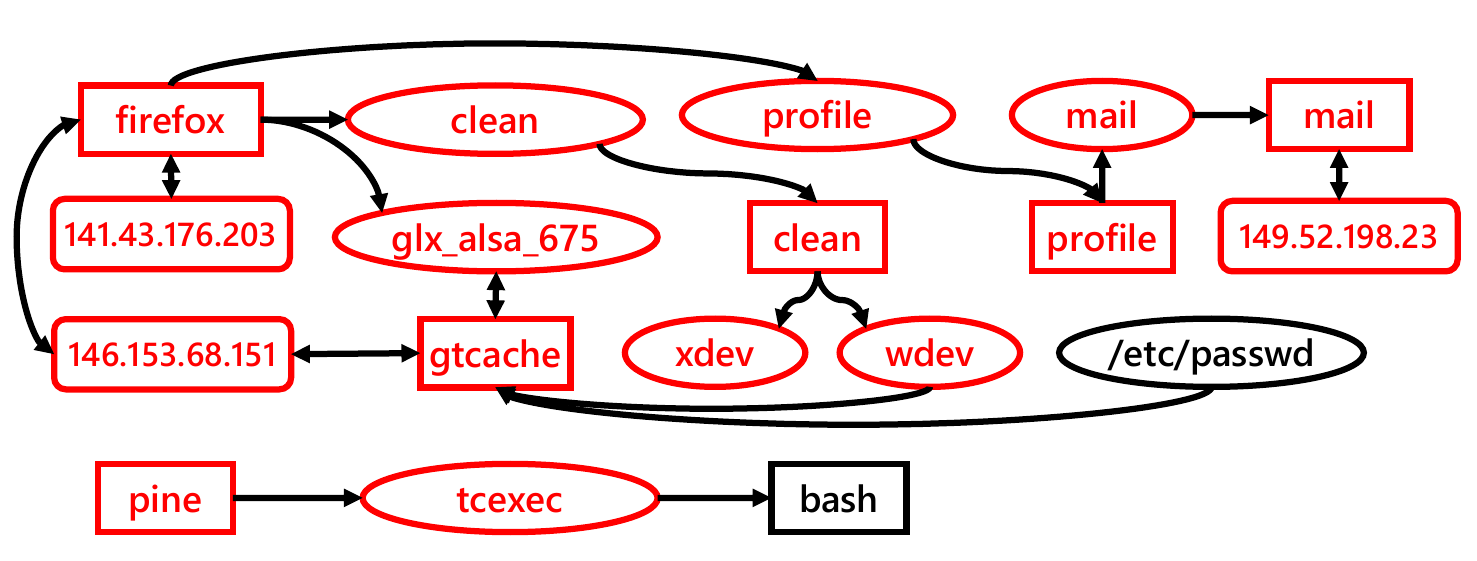}
    \caption{Attack graph of sub-dataset E3-THEIA.}
    \label{fig_theia_attack}
\end{figure}
\vspace{3pt}\noindent$\bullet$~E3-THEIA (Figure~\ref{fig_theia_attack}). Three successful attack attempts worked on THEIA: Firefox Backdoor, Browser Extension and Pine Backdoor. The user was compromised by a malicious payload \emph{clean} while browsing via \emph{firefox}. \textbf{clean} acquired root privileges, connected back to the attacker and executed another payload \emph{profile}. The Browser Extension attack aimed to resume the first attack by re-establishing the connection, grabbing root privileges and portscanning the local network via \textbf{gtcache} and \textbf{mail}. The Pine Backdoor attack is very similar to the one conducted on Trace but unexpectedly stopped due to missing library error.

\begin{figure}[t]
    \centering
    \includegraphics[width=0.48\textwidth]{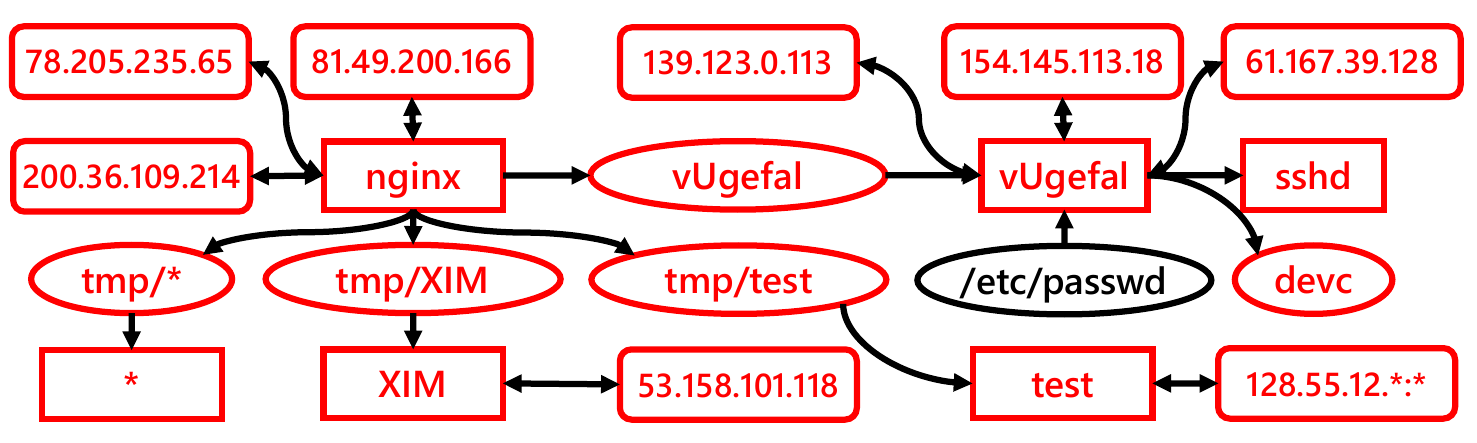}
    \caption{Attack graph of sub-dataset E3-CADETS.}
    \label{fig_cadets_attack}
\end{figure}
\vspace{3pt}\noindent$\bullet$~E3-CADETS (Figure~\ref{fig_cadets_attack}). The attacker exploited the Nginx Backdoor and tried two attacks on CADETS. During the first attack, the attacker connected to a vulnerable Nginx server running on CADETS and injected process \textbf{vUgefal} with root privileges. It read sensitive information and tried to further infect the \emph{sshd} process with malicious implants before the host crashed. The attacker then tried another attack on CADETS, resulting in a malicious process \textbf{XIM}. The attacker also created another process \textbf{test} to establish a long-lasting connection and portscan the local network.

\end{document}